\documentclass[epjST,final]{svjour}

\usepackage{graphicx,amsmath}

\usepackage{tabularx}
\newcolumntype{C}[1]{>{\centering\arraybackslash}p{#1}}

\begin{document}

\title{Transport properties of copper phthalocyanine based organic electronic devices}

\author{C.\ Schuster\thanks{\email{cosima.schuster@physik.uni-augsburg.de}}, M. Kraus, 
A. Opitz\thanks{\email{andreas.opitz@physik.uni-augsburg.de}}, W. Br\"utting, and U. Eckern}
\institute{Institut f\"ur Physik, Universit\"at Augsburg, 86135 Augsburg, Germany}
\date{\today}


\abstract{ 
Ambipolar charge carrier transport in Copper phthalocyanine (CuPc) is studied experimentally in
 field-effect transistors and 
metal-insulator-semi\-con\-ductor diodes at various temperatures. 
The electronic structure and the transport properties of CuPc attached to leads 
are calculated using density functional theory and scattering theory at the non-equilibrium 
Green's function level. 
We discuss, in particular, the electronic structure of CuPc molecules attached to gold chains 
in different geometries to mimic the different experimental setups.
The combined experimental and theoretical analysis explains the dependence of the mobility 
and the transmission coefficient on the charge carrier type (electrons or holes) and on the contact geometry.
We demonstrate the correspondence between our experimental results on thick films and 
our theoretical studies of single molecule contacts.
Preliminary results for fluorinated CuPc are discussed.}

\maketitle

\section{Introduction}

Charge carrier transport in electronic devices is an important topic, experimentally as well
as theoretically. Recently, theoretical progress has been made on different levels, 
for example, a parameter-free calculation of the transport 
behavior has been accomplished for metallic nano-contacts
and single-molecule devices \cite{smj,transiesta,smeagol}. It also has been possible to 
determine the transmission through interfaces, simple metals serving as leads, for 
example, Fe/MgO/Fe or Au/MgO/Au tunnel junctions \cite{rungger1,moh2}. In these studies,
the ground state properties of the system within a suitably defined central region, the
``scattering region'', are determined with the help of density functional 
theory (DFT) \cite{dft}. The electronic energy levels of the individual molecules,
monolayers and crystals in the scattering region largely determine the charge transport. 
Thereby a basic understanding of energy level alignment and the related injection 
barriers is achieved \cite{osc}. Finally, the transmission through 
the device is obtained by scattering theory at the non-equilibrium 
Green's function level \cite{Meir92}.

On the other hand, the temperature-dependent transport properties of organic thin-film 
devices can be modeled as two-dimensional drift-diffusion processes using commercial 
programs \cite{paasch}. In the standard device simulation software for hopping transport,
the possibility to treat a broad density of states, which would be relevant in the present 
context, is included. Of course, a combination of the classical stochastic treatment
of thin films with the quantum mechanical DFT approach for single-molecule contacts
would be very desirable \cite{lugli07}.

As a particular class of planar aromatic compounds, metal phthalocyanines are considered 
for numerous applications since they show a variety of interesting physical 
and chemical properties. 
In particular, the semiconducting copper phthalocyanine (CuPc, for the chemical structure 
see Fig.~\ref{figstruct}a) is employed in several optoelectronic devices \cite{shirota}.
It is used as buffer layer in organic light-emitting diodes \cite{bufferlayer} or as active 
layer in organic field-effect transistors (OFETs) \cite{active}. Due to the narrow $d$-bands of 
the transition metal and the $\pi$-bonding present on the benzene rings, CuPc crystals 
show effects which can only be accounted for by electronic correlations. The correlation 
effects manifest themselves, for example, in the Mott metal-insulator transition upon 
K-doping \cite{craciun,MIT_th}. However, in the following we will concentrate on certain
aspects of transport, i.e., charge injection and the mobility of charge carriers,
since these are of practical interest for applications in optoelectronic devices.

In this paper we present the results of a combined experimental and theoretical study. 
The experimental results are obtained by measuring field-effect devices, while for the 
theoretical part we employ DFT and the non-equilibrium Green's function formalism. In both
parts the focus is on the role of the contact geometries.  
In the following Secs.~2 and 3, we describe the experimental and the theoretical details, respectively. In 
Sec.~4 we focus on ambipolar field-effect devices; we find that the experimental characteristics
can be well explained on the basis of DFT. Our results for the temperature dependence of the
electron and hole mobilities are presented and discussed in Sec.~5. In Sec.~6 we argue
that the mobility experiments can be understood by studying a simple model system, namely,
a single CuPc molecule, or two CuPc molecules, attached to Au chains which serve as leads. 
Finally, we discuss in Sec.~7 some aspects of fluorinated CuPc, and present a brief
summary in Sec.~8.

\section{Experimental details}

\begin{figure}[b]
\centering{\includegraphics[width=.85\linewidth]{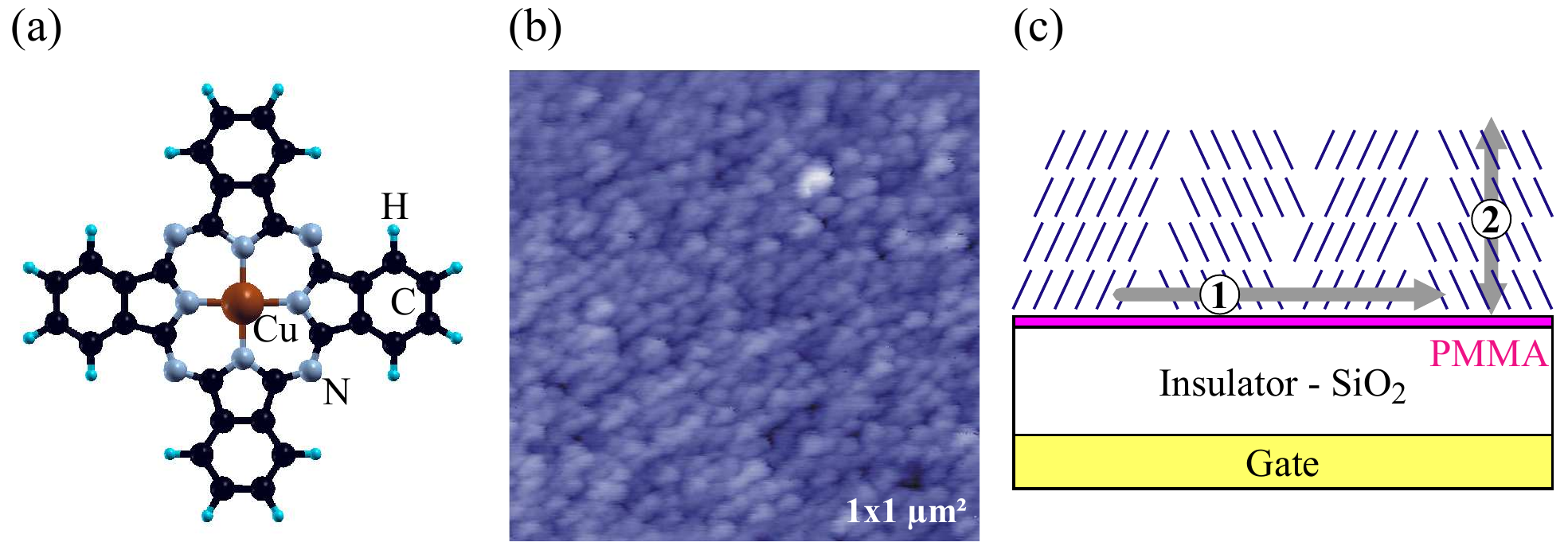}}
\caption{(a) Molecular structure of CuPc. (b) Scanning force microscopy image of an evaporated 
CuPc film deposited on a PMMA passivation layer. The height scale is about 9\,nm from black 
via blue to white.
(c) Schematic structure of the transport in field-effect devices. The gray arrows 
illustrate the charge transport directions in the devices. Direction (1) represents the transport 
in a field-effect transistor (FET), and direction (2) the transport in a
metal-insulator-semiconductor (MIS) diode. The top contact 
is omitted for clarity. The molecules are standing almost upright on this substrate 
surface \cite{Hos03}. The observed grain boundaries are illustrated.}
\label{figstruct}
\end{figure}

CuPc -- the molecular structure is shown in Fig.~\ref{figstruct}a -- 
has been purchased from Sigma Aldrich as sublimation grade and was additionally purified 
by temperature gradient sublimation. The material was evaporated at a pressure of 
$1.0\times 10^{-7}$\,mbar at a rate of 0.2\,\r{A}/s. The resulting film morphology 
determined by scanning force microscopy in non-contact mode is shown in Fig.~\ref{figstruct}b. 
A typical polycrystalline structure is observed with an average crystal diameter of 
about 40\,nm.  

Field-effect devices were fabricated on highly doped $p^{++}$-Si wafers with 320\,nm thermally 
grown SiO$_{2}$ acting as gate electrode and gate dielectric, respectively. A schematic sketch 
of the OFETs can be seen in Fig.~\ref{figstruct}c. The wafers have been purchased from Si-Mat, 
Germany. Cleaning of the wafers was performed in an ultrasonic bath with acetone and isopropanol 
successively.  After cleaning, PMMA was spin-coated as a passivation layer from a 1.0\,wt.\% 
solution in toluene on top of $\mathrm{SiO_2}$. The active semiconductor layer consists of a 
25\,nm thick layer of CuPc deposited on top of the passivated substrates for FET measurements 
and for metal-insulator-semiconductor
(MIS) diodes. For the transistors, top source and drain contacts were evaporated through 
a shadow mask with various channel lengths in the range of 80\,$\mu$m to 180\,$\mu$m using 
different metal contacts with a thickness of 50\,nm. In some cases a thin (approximately 1\,nm) 
interlayer of the strong organic acceptor F$_4$TCNQ is evaporated between the organic semiconductor and the 
metal top contact. The areas of the MIS diodes 
were about 5.0\,mm$^2$, defined by the structure of the semiconductor and 
the top contact. Transistor 
characteristics were measured with a Keithley 4200 semiconductor parameter analyzer, and for 
the impedance spectroscopy a Solartron 1260 impedance/gain-phase analyzer coupled with a 
Solartron 1296 dielectric interface was used.

In the case of OFETs with different electrode materials the charge carrier mobility
was determined
using the so-called transmission line method (TLM) \cite{luan92}. This technique allows to 
extract the charge carrier mobility $\mu_{\mathrm{TLM}}$ without the disturbing 
effect of the contact 
resistance. In most cases the electrode-semiconductor interface is not ideally ohmic but 
forms a Schottky contact which causes an injection barrier depending on the relative positions of 
the HOMO and the LUMO of the semiconductor and the work function of the electrode metal. The 
total resistance of the OFET $R_{\mathrm{total}}$ is split into the contact resistance 
$R_{\mathrm{C}}$ (which is independent of the channel length) and the channel resistance 
which is proportional to the channel length $L$. The total resistance is then given by
\begin{equation}
\label{eq1}
R_{\mathrm{total}}
= R_{\mathrm{C}}+\frac{L}{\mu_{\mathrm{TLM}}WC_{\mathrm{ins}}(V_{\mathrm{G}}-V_{\mathrm{T}})}
\end{equation}
with the channel width $W$, the specific insulator capacitance $C_{\mathrm{ins}}$, the gate 
voltage $V_{\mathrm{G}}$ and the threshold voltage $V_{\mathrm{T}}$.
Thus, a plot of the total resistance as a function of the channel length allows to determine
$\mu_{\mathrm{TLM}}$ (slope of the linear fit).

In case of temperature dependent measurements the field-effect mobility is calculated 
from  the transfer characteristics in the linear regime, using the following expression 
for the drain current $I_{\mathrm{D}}$
\begin{equation}
\label{eq2}
I_\mathrm{D}
=\frac{WC_{\mathrm{ins}}}{L}\mu_{\mathrm{lin}}(V_{\mathrm{G}}-V_{\mathrm{T}})V_{\mathrm{D}}
\end{equation}
where $V_{\mathrm{D}}$ is the drain voltage. Clearly the mobility 
$\mu_{\mathrm{lin}}$ can be derived from the slope of the transfer characteristic,
$I_{\mathrm{D}}$ versus $V_{\mathrm{G}}$.

In contrast to OFETs where the charge carrier mobility parallel to the semiconductor-insulator 
interface is measured, in MIS diodes the charge carrier mobility perpendicular to this 
interface can be determined. Due to the structural anisotropy of  organic semiconductor molecules 
the overlap of the electronic orbitals and thus the charge carrier transport is completely 
different in these two directions. 

The charge carrier mobility perpendicular to the surface, $\mu_{\rm MIS}$, can be determined
with the help of impedance spectroscopy. The capacitance of the MIS diode 
is measured as a function of the applied frequency in the accumulation regime of electrons or
holes, i.e., at positive or negative gate voltage, respectively. For low frequencies, due to the 
accumulation of charge carriers at the interface, only the 
capacitance of the insulating layer $C_{\mathrm{ins}}$ (i.e., $\mathrm{SiO_2}$ and PMMA) is 
measured. When the frequency is 
increased above a characteristic frequency $f_c$ the charge carriers cannot follow the external 
field any more because of the limited mobility  in the CuPc layer. Consequently, at $f_c$ 
the measured capacitance drops to the geometric capacitance of the whole device, 
$C_{\mathrm{MIS}}^{-1}=C_{\mathrm{ins}}^{-1}+C_{\mathrm{CuPc}}^{-1}$. The motion of the 
charge carriers can be assumed to be determined by diffusion \cite{stallinga08}. Following this 
approach, the mobility is given by
\begin{equation}
\label{eq3}
\mu_{\rm MIS}=\frac{2\pi f_{\mathrm{c}} e d_{\mathrm{s}}^2}{kT}
\end{equation}
with the elementary charge $e$, the thickness of the semiconductor $d_{\mathrm{s}}$, the 
Boltzmann constant $k$ and the temperature $T$.

\section{Computational details}
In the following, we analyze the electronic structure, in particular, the charge density 
of the HOMO (highest occupied molecular orbital) and the LUMO (lowest unoccupied molecular orbital) 
using density functional theory. In particular, we employ the DFT package SIESTA \cite{Sol}, 
which relies on a basis set of local atomic orbitals. In addition, SIESTA uses norm-conserving 
pseudo-potentials in the fully non-local form
(Kleinman-Bylander \cite{kleinman}). For the metals Cu and Au, we apply pseudo-potentials
including $d$ valence states.
Moreover, a double zeta basis set and the generalized gradient approximation (GGA) for the
exchange correlation potential are used.

As CuPc contains a transition metal within porphyrin and benzene rings, interaction effects are expected
to be relevant, and we have to choose the ``best possible'' exchange correlation potential 
for the present purpose. In particular, spin polarization is an important ingredient, since
Cu$^{2+}$ shows a significant spin splitting \cite{bialek}.
In a systematic study, Marom et al.\ calculated the electronic structure of CuPc using 
 LDA (local density approximation) and GGA, 
as well as hybrid functionals like the semi-empirical B3LYP, the 
non-empirical PBE0, and the screened SHE03 \cite{marom08}. They concluded that for
CuPc/metal systems the screened hybrid functional is the best choice,
since it performs very well for CuPc and is reasonable also for metals \cite{paier07}. In
another study, the B3LYP functional was used to determine and classify molecular orbitals 
\cite {mertig}. 
However, we are predominantly
interested in transport properties, i.e., the most important states near the Fermi level
which are sufficiently well described in GGA \cite{marom08}; also GGA is the best choice 
for noble metals \cite{paier07}. Hence we rely on the GGA in the following.

\begin{figure}
\includegraphics[width=0.55\textwidth]{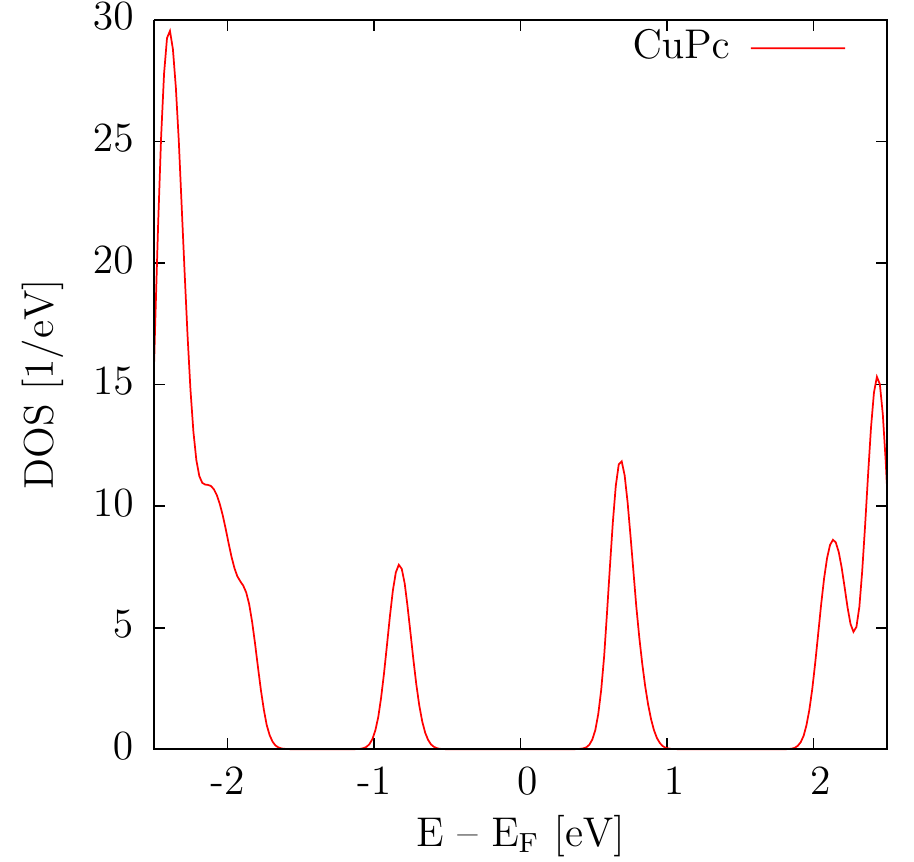}
\caption{Density of states of CuPc in a spin polarized GGA calculation, for different
substrate configurations as explained in the main text. A level broadening of 0.1 eV 
was used for the plot.}
\label{fig_DOS}
\end{figure}

Before turning to CuPc molecules in contact with a metal, we summarize 
the results obtained for CuPc monolayers. Monolayers of flat lying molecules 
 are formed on metallic surfaces or graphite;
the typical molecule-molecule distances in a CuPc monolayer on top of a [111] directed 
surface of Au \cite{CuPcAu} or a [110] surface of Ag \cite{AgCuPc}
are larger than the molecule-molecule distances in the CuPc crystal \cite{cupcp}.
The CuPc molecules in a monolayer show different orientations towards each other, depending on
the symmetry of the substrate. However, due to the large molecule-molecule distances within 
the monolayer the electronic structure of CuPc near the Fermi level is very similar to that
of a single molecule, with only minor effects of the coordination to the other molecules.
We have checked this supposition for an orientation of the molecules to each other of 60$^\circ$ 
and a molecule-molecule distance of $d=14.4$~\r{A}\ (as obtained for a Au[111] substrate),
a 90$^\circ$ orientation with $d=17.35$ \r{A}\ (Ag[110] substrate), and a 45$^\circ$ 
orientation with $d=17.35$ \r{A}. 
The density of states (DOS) near the Fermi level, shown in Fig.~\ref{fig_DOS},
is identical in all these configurations. In addition, it agrees with recent 
theoretical and experimental results \cite{bialek,alvarado}. We also
studied a 45$^\circ$ configuration with a smaller distance, $d=13.72$ \r{A}, 
corresponding to the bulk value.
In this case, the band gap is slightly smaller, 1.43 eV versus 1.52 eV, than the 
HOMO-LUMO gap of single molecules, in accordance with previous results \cite{mertig}.
The calculation thus  underestimates (as to be expected in LDA and GGA)
the experimental gap of 1.8 eV \cite{kraus08}.
Almost identical photoemission spectra of CuPc thin films 
\cite{mertig} and gase phase CuPc \cite{zitatausmertig} also
suggest that the electronic structure is hardly modified by inter-molecular interactions.

\begin{figure}
\includegraphics[width=0.8\textwidth]{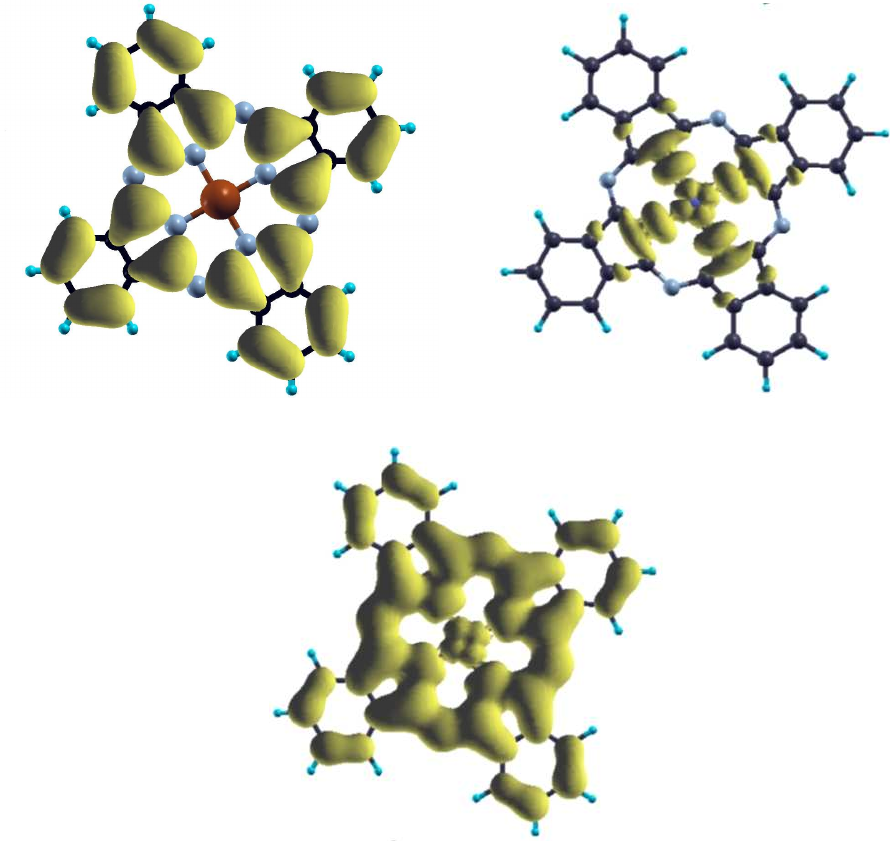}
\caption{Charge density isosurfaces of relevant orbitals. Top: the HOMO contains contributions
from the a$_{1u}$ orbital (left hand side) and from the b$_{1g}$ orbital (right hand side) located 
on copper. Bottom: the LUMO contains contributions from of the e$_{g}$ orbital and from 
the b$_{1g}$ orbital; here, the  sum of both contributions is shown.}
\label{fig-mo}
\end{figure}

For the discussion of the transport properties we analyze the molecular orbitals near to 
the Fermi level.
In particular, the HOMO includes a contribution of a (single occupied) b$_{1g}$ orbital localized 
on the Cu together with the a$_{1u}$ molecular orbital, and the LUMO a b$_{1g}$ 
orbital together with the e$_g$ molecular orbital. 
The charge density isosurface is obtained by integrating the DOS, $N(E)$, in a certain energy range,
$n({\bf r})=\int_{E_1}^{E_2} {\rm d}E \, N(E)$.  
A plot of the charge density isosurface of the 
states contributing to the HOMO ($E_1=-1$ eV, $E_2=-0.65$ eV) and the LUMO ($E_1=0.5$ eV, $E_2=1$ eV) 
is shown in Fig.~\ref{fig-mo}.
Our results for the  charge density of the LUMO are very similar to those obtained in  \cite{mertig}.
In case of the HOMO, the Cu b$_{1g}$ contributes to the charge density within our calculations (using GGA)
 but not within the previous studies using B3LYP \cite{mertig}.
The most important difference of the HOMO versus the LUMO lies in the missing contribution 
on the nitrogen in the HOMO. Otherwise the spatial extensions are almost identical, and hence
do not allow conclusions concerning the 
overlap of the molecules.

\section{Ambipolar field-effect devices}

\begin{figure}[b]
\centering{\includegraphics[width=.45\linewidth]{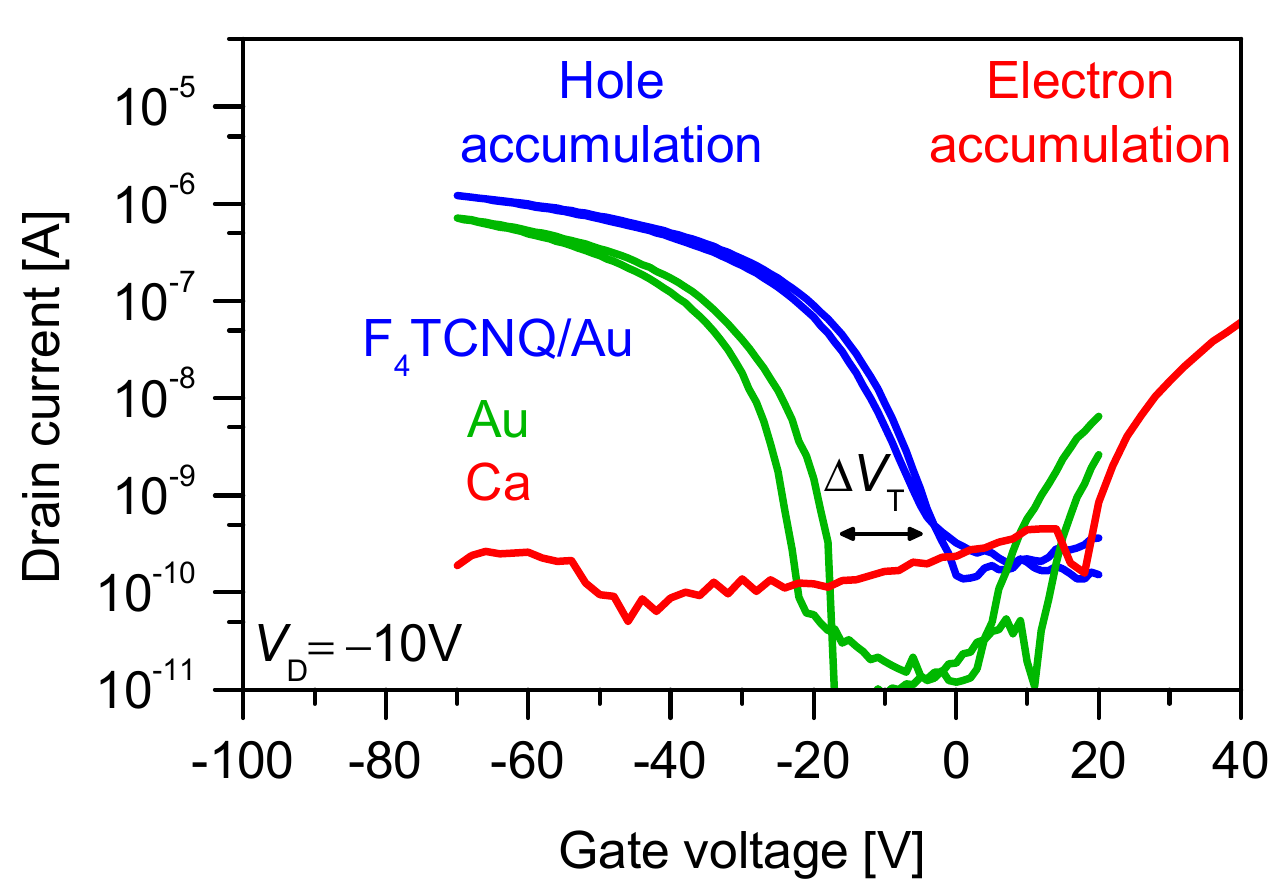}\hfill
\includegraphics[width=.45\linewidth]{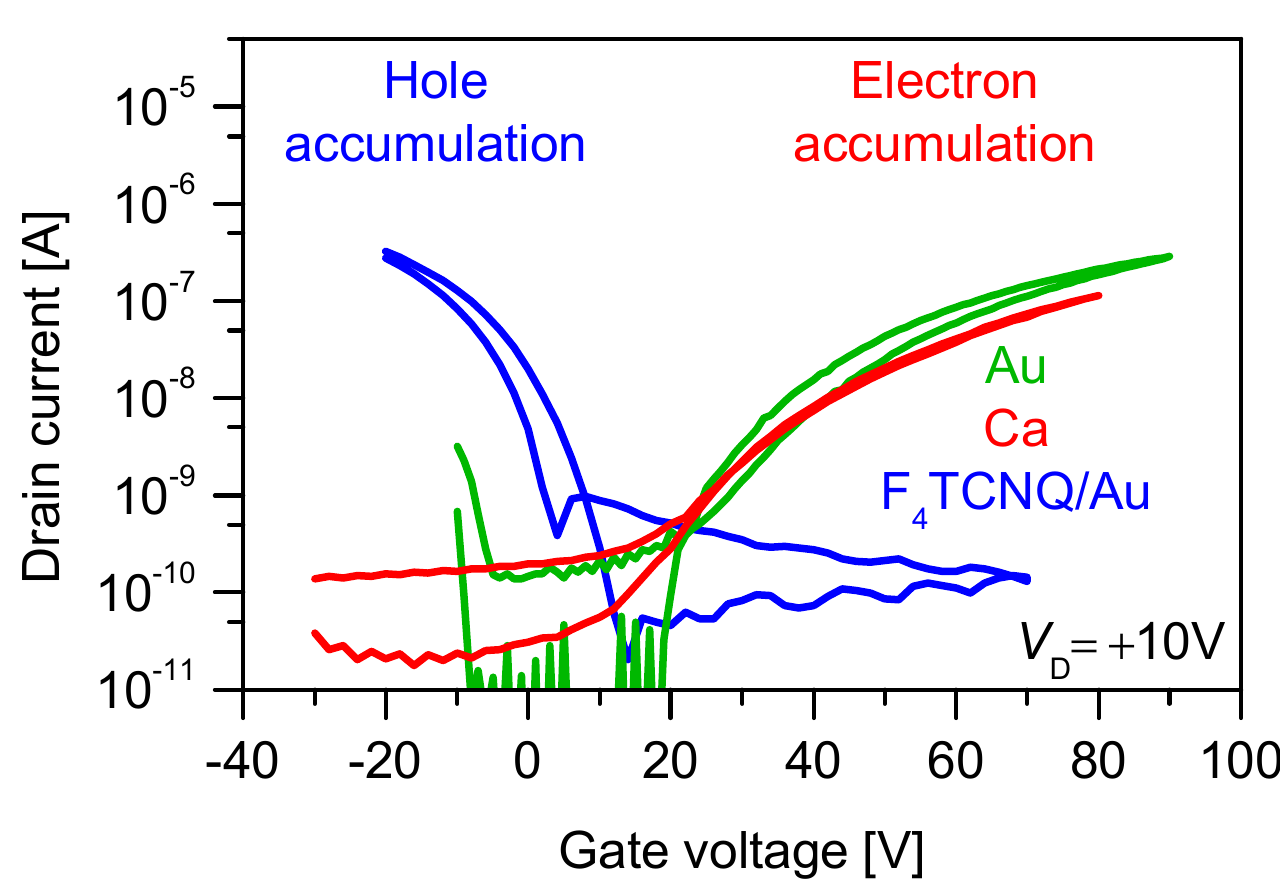}}
\caption{Transfer characteristics of CuPc FETs in the top-contact configuration with 
different electrodes: calcium, gold, and F$_4$TCNQ/gold. The threshold voltage shift 
$\Delta V_\mathrm{T}$ between the green and the blue curve is related to a coverage of the whole semiconducting film with 
F$_4$TCNQ \cite{dvt}.}
\label{figFET}
\end{figure}

\begin{figure}[b]
\centering{\includegraphics[width=.45\linewidth]{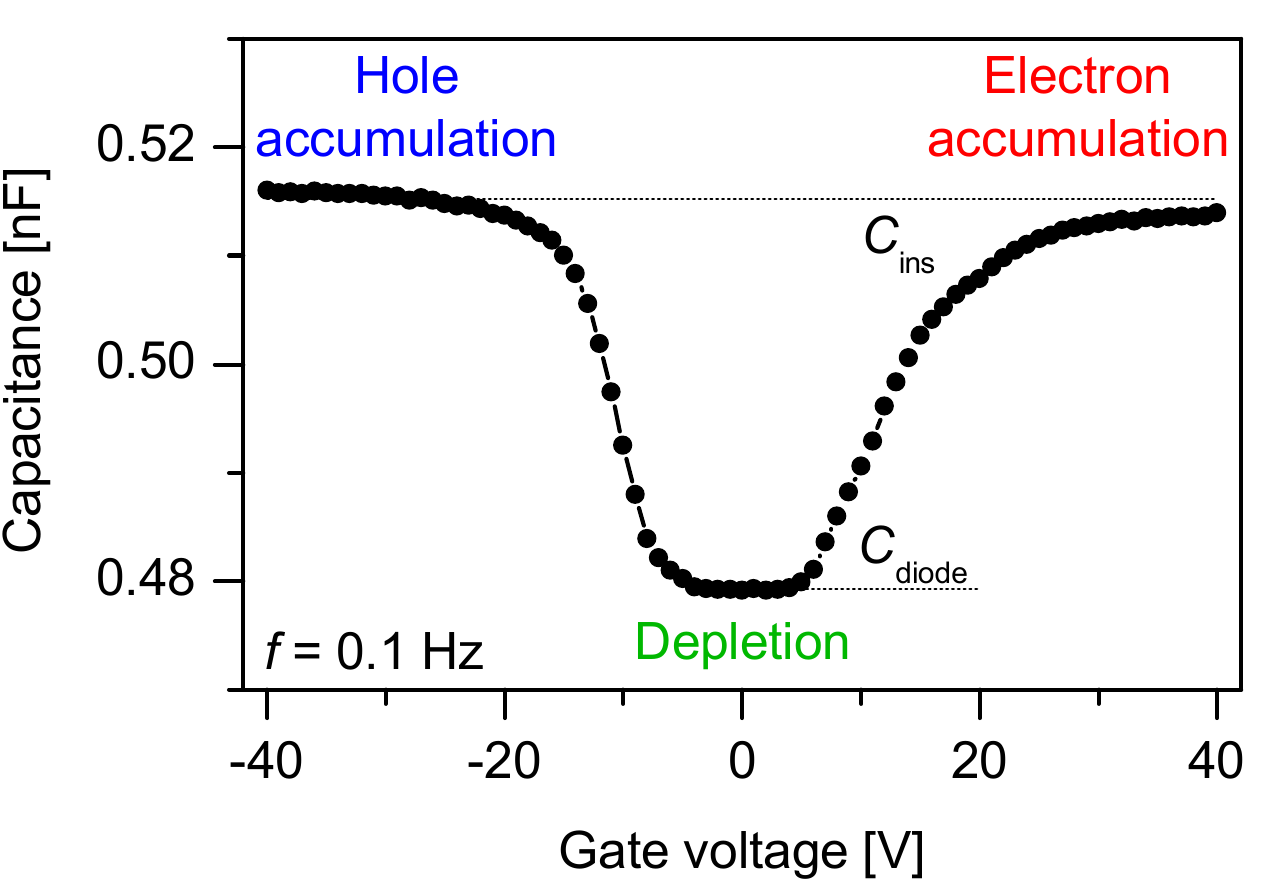}\hfill
\includegraphics[width=.45\linewidth]{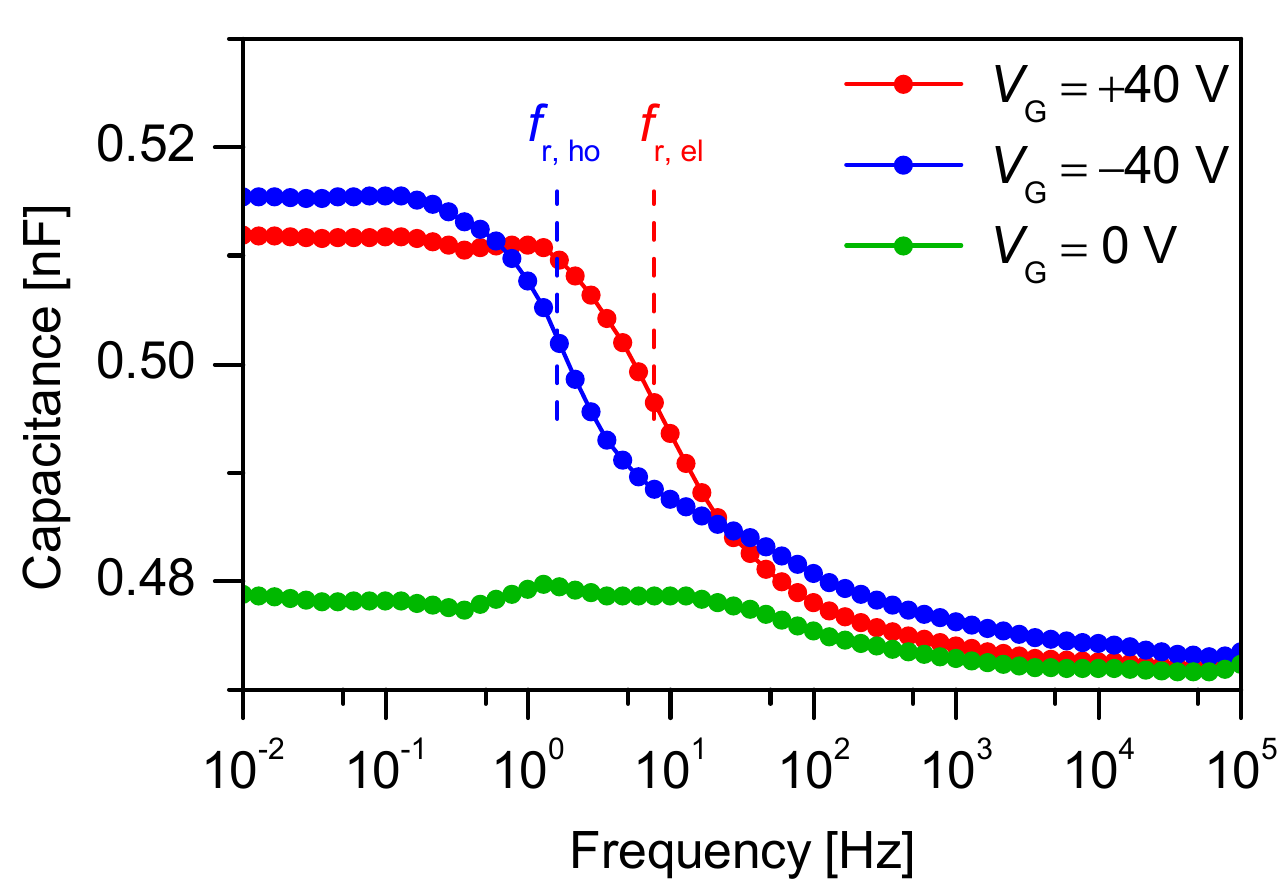}}
\caption{Capacitance spectroscopy measurements (left: capacitance versus voltage, right: 
capacitance versus frequency) of a CuPc MIS diode with gold top contacts.}
\label{figMIS}
\end{figure}

\begin{table}[t]
\renewcommand{\arraystretch}{1.2}
  \centering
  \caption{Room temperature mobilities for electrons and holes in FETs using different 
   electrode materials, and in MIS diodes using gold contacts.}
  \label{tabmob}
\begin{tabular}{C{0.2\linewidth}C{0.2\linewidth}C{0.2\linewidth}}
\hline
Devices & Electron transport & Hole transport\\\hline\hline
&\multicolumn{2}{c}{Mobilities in FETs [cm$^2$/Vs]}\\ \hline
Ca contacts & $3 \times 10^{-4}$ & -- \\
Au contacts & $5 \times 10^{-4}$ & $2 \times 10^{-3}$\\
F$_4$CNQ/Au contacts & -- & $3 \times 10^{-3}$\\\hline
&\multicolumn{2}{c}{Mobilities in MIS diodes [cm$^2$/Vs]}\\ \hline
Au contacts & $3 \times 10^{-8}$  & $1 \times 10^{-8}$\\\hline
\end{tabular}
\end{table}

Ambipolar charge carrier transport in field-effect devices is related to the 
accumulation of both charge carrier types, electrons and holes, at the semiconductor/insulator 
interface. To prevent traps at the oxide surface a polymeric insulator is used in this study 
as passivation layer \cite{general-n,opitz08}. For both device types, FET and MIS diode, electron and hole 
accumulation are observed by using gold electrodes.
However, as shown in Figs.~\ref{figFET} and \ref{figMIS},
 the accumulation of charge carriers depends on the applied gate voltage. While 
electrons can be accumulated by applying a positive gate voltage, holes will be accumulated 
by applying a negative one. The accumulation of the respective charge carriers leads to 
an increasing drain current in an FET and to a capacitance comparable to the insulator capacitance in an 
MIS diode. Depending on the gate voltage the switch-on voltage (FET) or the flatband 
voltage (MIS diode) can be defined as voltage where the accumulation starts. Between the 
onset voltages for accumulation of holes and electrons the region of depletion is measured. 
There the current of an FET is in the noise level, and the capacitance of the MIS diode is 
defined by a series circuit of the insulator and the semiconductor capacitance.

The charge carrier type can also be controlled by varying the electrode material 
(see Fig.~\ref{figFET}). By using Ca electrodes instead of Au the injection of electrons is 
enhanced due to the low work function of the electrode material, and unipolar electron transport 
can be detected. In contrast, by using an F$_4$TCNQ interlayer the injection of electrons can be 
suppressed and unipolar hole transport is observed. Using TLM, the charge charier mobilities 
are determined as described in connection with Eq.~(1); see Table~\ref{tabmob}. 
For CuPc in the FET configuration 
the mobility of holes is almost one order of magnitude higher than the mobility of electrons.
The analysis for different electrodes shows that the mobilities are 
independent of the electrode material \cite{opitz08}. 

The transition from accumulation to depletion in the MIS diode can be seen in capacitance 
spectroscopy using capacitance-voltage ($C$-$V$) and capacitance-frequency ($C$-$f$)
measurements. The mobilities of the ambipolar MIS diode are 
also given in Table~\ref{tabmob}. The MIS values  are four to five 
orders of magnitude lower than the mobilities in the FETs, while the respective 
differences of the electron and the hole mobilities are less pronounced.

\begin{figure}
\centering{\includegraphics[width=0.45\textwidth]{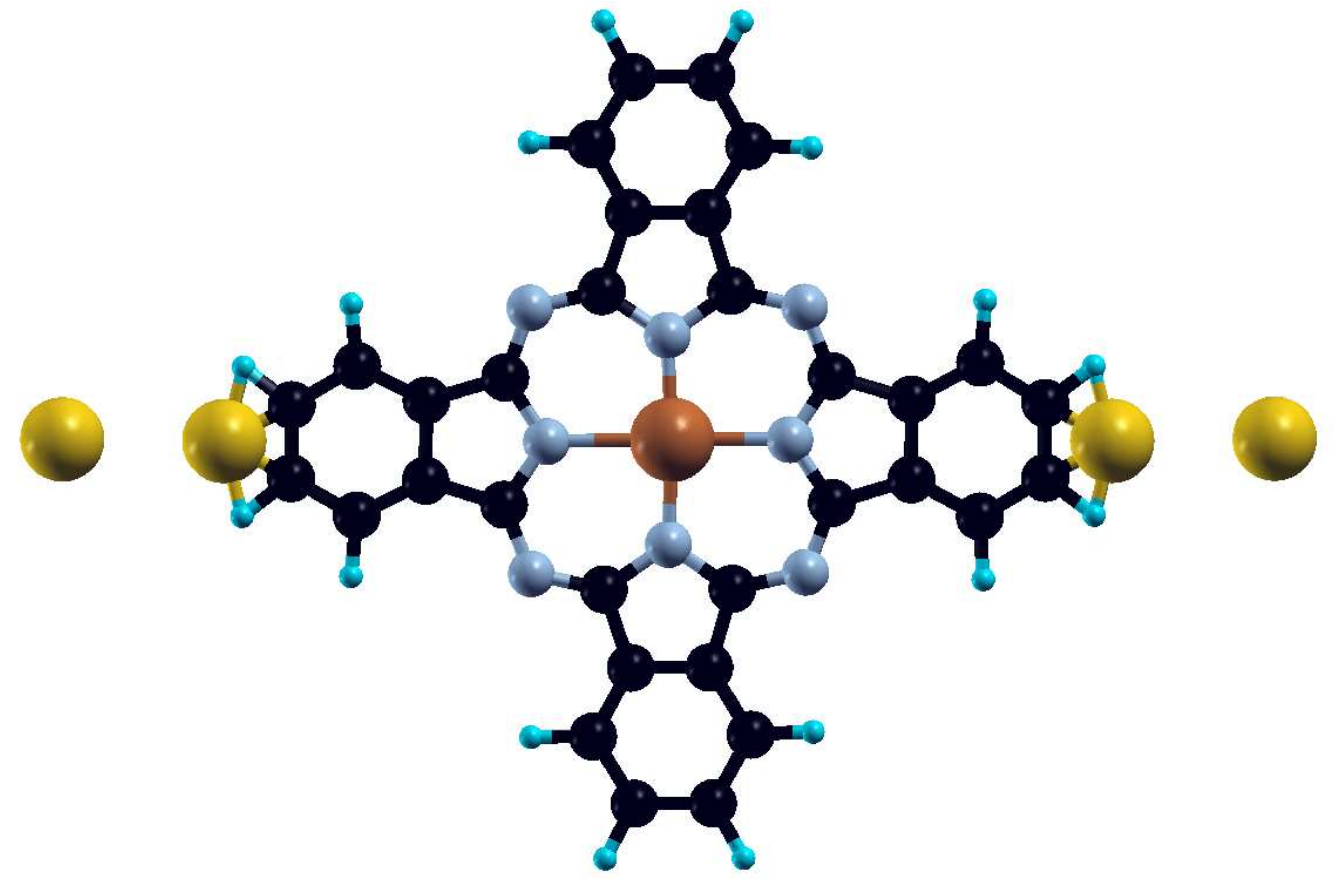}\hspace*{1cm}
\includegraphics[width=0.3\textwidth,angle=90]{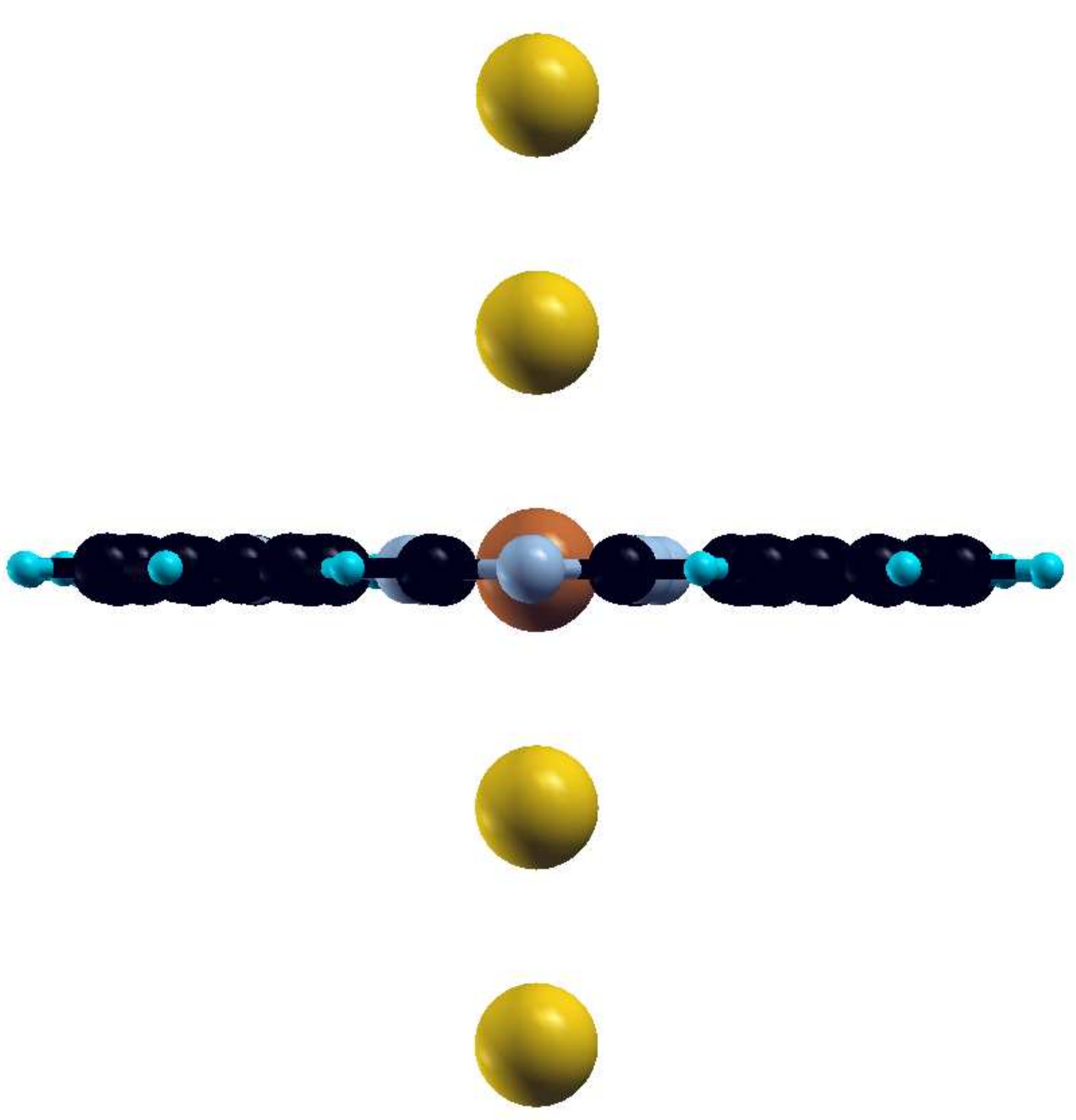}}
\caption{Left: Top view of the planar contact. The distances Au-H and Au-C are given 
  by $d_{\rm Au-H/C}=1.37$ \r{A}.
  Right: Sideview of the perpendicular contact. The distance Au-Cu is given by 
  $d_{\rm Cu-Au}=2.89$ \r{A}.}
\label{CuPc_flach}
\end{figure}

Our DFT results for a single molecule attached to Au chains show that
the electronic structure near the Fermi level of a contacted molecule 
is similar to that of an isolated one.
Note that in this calculation the contact of the Au chains to the molecule is established via
the outer phenyl rings in a planar transport geometry, see Fig.~\ref{CuPc_flach}, 
left hand side. (Another contact configuration will be discussed in Sec.~6.2.)

Experimentally, this configuration -- a  prototypical
onedimensional system -- was realized by placing the CuPc molecule and the Au atoms with 
a scanning tunneling microscope (STM) tip on a NiAl substrate \cite{nazin03}. 
The distances between Au and H, and between Au and C, are obtained by strucutural 
relaxation;  both are equal to $d=1.37$ \r{A}. 
The distance between the Au atoms in the chains is fixed to the bulk nearest-neighbor distance of $2.89$ \r{A}.
As Ca does not form chains in nature (and also because the DFT for 
Ca chains does not converge to a self-consistent solution),
we contact the CuPc molecule to bulk Ca in the [001] fcc 
direction using a pyramidal contact geometry. Using the planar configuration we can now 
compare contacts made of different materials -- Au chains versus bulk Ca contacts --
to determine the influence of the material and the dimensionality of the leads.
Recently, it was shown in fact that an insulating six-site Au chain attached to Au chains as compared to
pyramidal leads has similar transport properties \cite{ThomDiplom}. In this model study,
it was found that due to the different coupling
the molecular levels are shifted by approximatively 0.1 eV in energy, and that
most of the differences between chain and
bulk contacts concern the Au $d$ states in the scattering region.

\begin{figure}
\begin{center}
\includegraphics[width=0.6\textwidth]{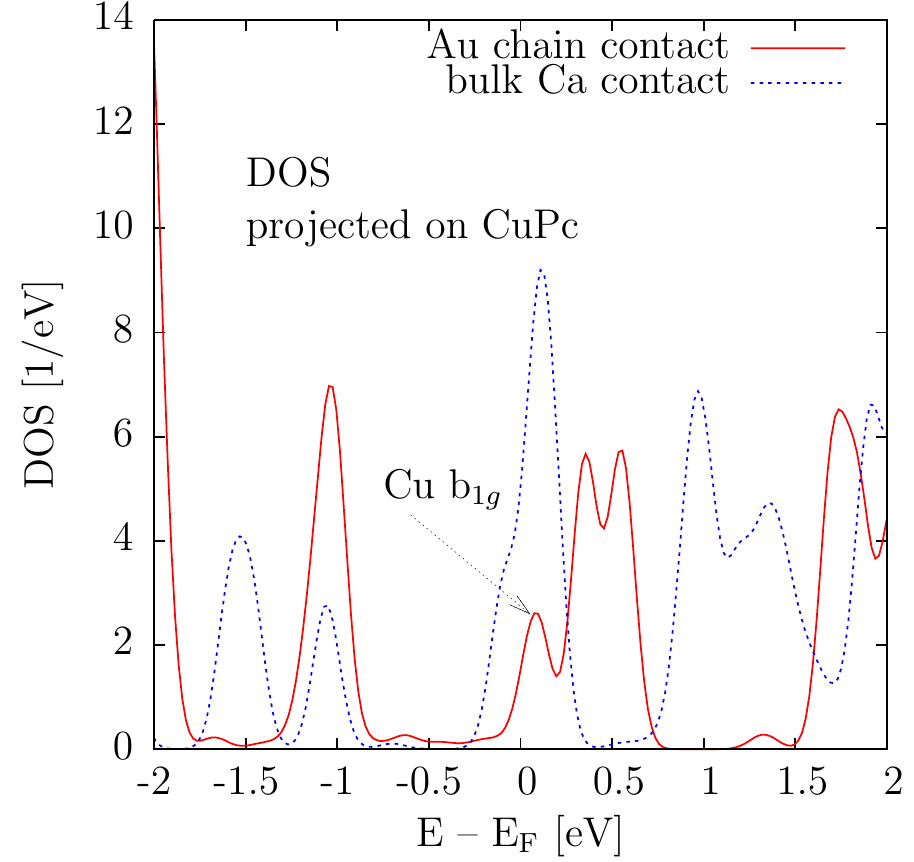}
\end{center}
\caption{DOS for CuPc contacted with Au chains or Ca bulk leads.}
\label{dos_au_ca}
\end{figure}

\begin{figure}
\centering{\includegraphics[width=0.9\textwidth]{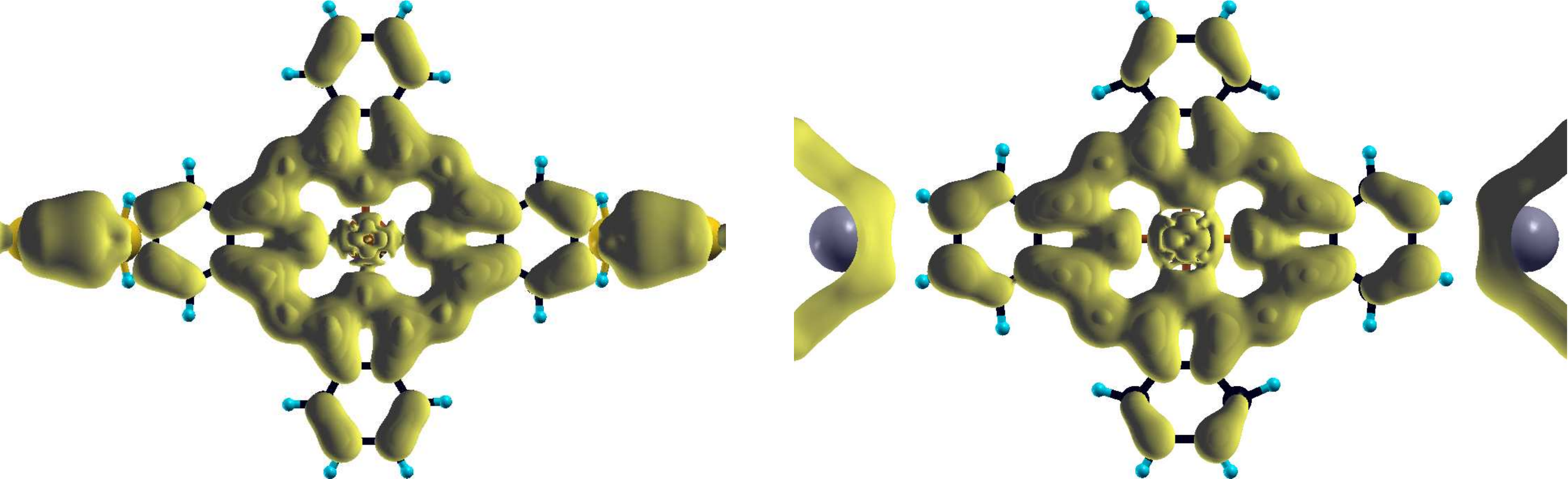}}
\caption{Left: Charge density isosurface at the Fermi level for 
CuPc contacted with Au chains.
Right: same, but CuPc contacted with bulk Ca. In both cases, electrons are injected onto 
the molecule.}
\label{cupc_mo}
\end{figure}

Concerning the electronic structure of the planar contact,
the DOS projected onto the CuPc states shows contributions at the Fermi level
in both cases, i.e., for Au and Ca contacts, see Fig.~\ref{dos_au_ca}.
In addition, the charge density isosurface of the states between $-0.3$ and 0.7 eV 
in the case of Au contacts,
and the charge density isosurface of the states between $-0.3$ and 0.4 eV in the case of 
Ca contacts, are almost identical, see Fig.~\ref{cupc_mo}.
For this reason, we  expect similar transport properties for Au and Ca contacted molecules.
Moreover, the shape of the molecular orbital
near E$_{\rm F}$ resembles the LUMO of the free molecule, indicating an electron transfer to 
the molecule due to the contacts.
Analyzing the contributions at the Fermi energy of the Au contacted molecule in detail,
we find that the Cu b$_{1g}$ dominates between $-0.3$ and 0.3 eV, whereas the e$_{g}$
dominates around 0.5 eV. Thus there is no overlap between
the electronic states of the leads and of the molecule near E$_{\rm F}$.
Since the shape of the molecular orbitals is not influenced by the leads,
the transport properties can be explained on the basis of the properties of an isolated molecule.


It is remarkable that samples with Au electrodes show ambipolar charge carrier transport 
although the work function of Au fits very well to the HOMO of CuPc \cite{opitz08}. This 
effect can probably be attributed to the top contacts. However, the work function of the metal and the HOMO level
cannot be directly compared because of the formation of
interface dipoles at the metal-organic interface that disrupt the
vacuum level alignment \cite{Yan01}. Additionally diffusion of the
metal into the organic film at high deposition temperature \cite{diff}
and deposition induced defects at the semiconductor/metal
interface \cite{topcontact} have to be considered. Our results indicate
that the effective work function of the top contact Au electrode allows injection of
both charge carrier types, and that it differs from
the work function of a bottom contact electrode.

As far as single molecules are concerned, we observe electron injection into the molecule
when the contact geometry is planar. In order to study  hole injection, we thus have to investigate other 
systems.  As a first attempt, we studied another electrode material, namely Pt since
 it  has the largest work function among the metals used as contacts. 
However, for this case  we find no overlap of the contact states with the molecular states, hence no contributions 
at the Fermi level and thus an insulating behavior.

\begin{figure}
\includegraphics[width=0.6\textwidth]{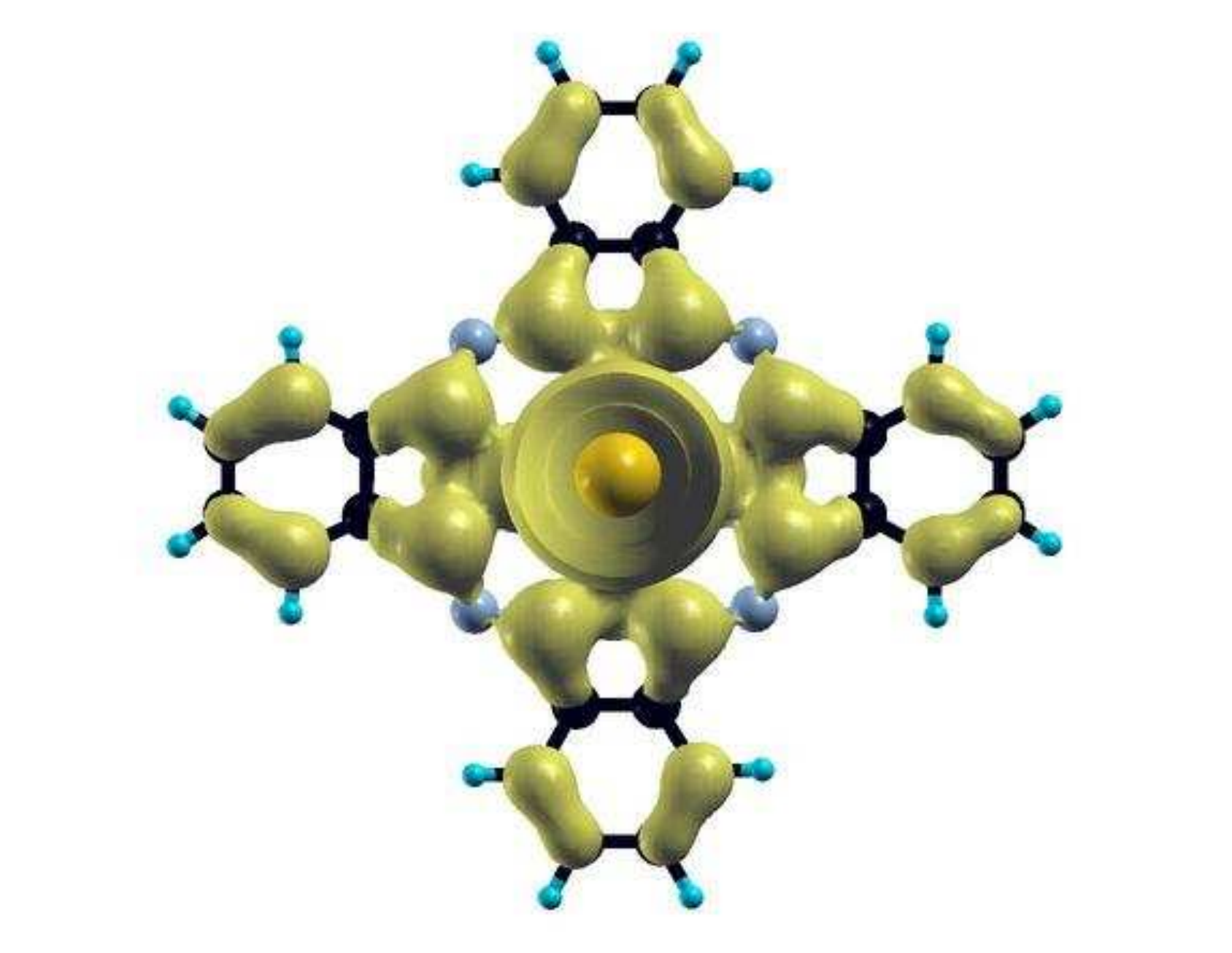}
\caption{Charge density isosurface at the Fermi level for CuPc contacted 
   with Au chains in a perpendicular geometry.}
\label{cupc_flach_mo}
\end{figure}

In addition, a different contact geometry was investigated.
We use again Au chains as leads but contact the molecule in a perpendicular geometry, 
see Fig.~\ref{CuPc_flach}, right hand side. Thus
we contact directly to Cu, simulating a transport measurement along path 1 
in Fig.~\ref{figstruct}c. The distance Au-Cu is chosen to be
equal to the Au-Au distance within the chains, $d_{\rm Cu-Au}=2.89$ \r{A}; for this value,
the force between Au and Cu is found to be extremely small.
Indeed, injection of holes, i.e., a charge density isosurface near the Fermi level resembling the 
HOMO of the isolated molecule, is found in the perpendicular configuration, see Fig.~\ref{cupc_flach_mo}.
In contrast to Au, Ca is expected to have no overlap with the Cu atom of the molecule, 
therefore the hole injection from Ca into CuPc in a perpendicular geometry should be blocked. 
Similarly, we have shown recently for bulk Au that metallic impurites with an electronic structure 
different from Au overlap only slightly with the host \cite{Moh3} and thus block the transport.

To summarize briefly, on the basis of DFT for single molecules in contact with a metal we have 
shown that electrons are injected in a planar contact, independent of the lead material,
whereas holes are injected in the perpendicular configuration  with  Au as electrode material. 
The contact resistance and the mobility can be obtained from the transmission 
coefficient, as discussed in Sec.~6.

\section{Charge carrier mobility}

\begin{figure}[b]
\centering{\includegraphics[width=0.6\linewidth]{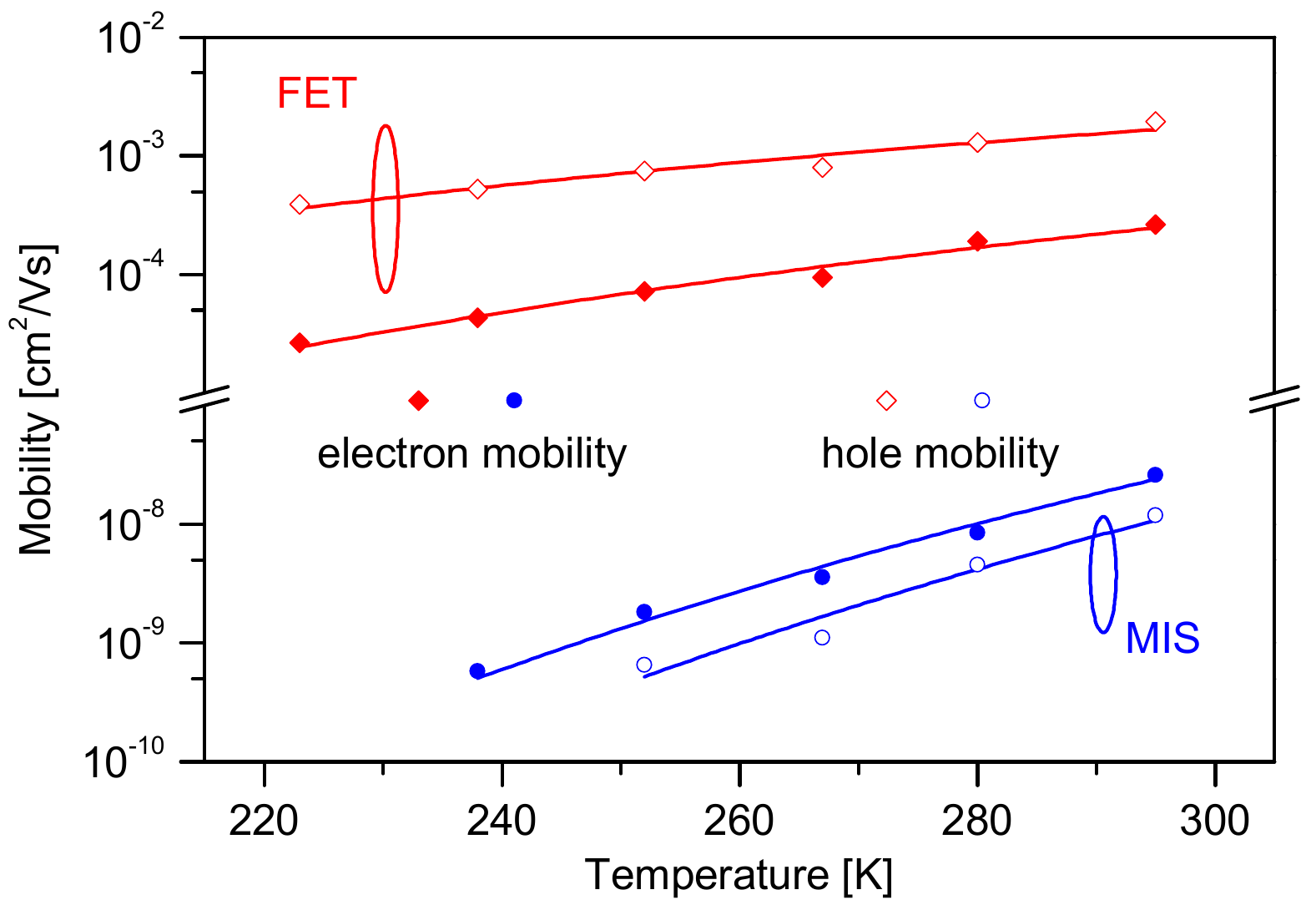}}
\caption{Temperature dependent mobilities for FET and MIS diode, and for 
  both charge carrier types. The lines are fits to determine the activation energy 
  using Eq.~(\ref{eq_ae}).}
\label{figmu}
\end{figure}

The temperature dependence of electron and hole mobilities for FETs and MIS diodes is shown 
in Fig.~\ref{figmu}. The hole mobilities determined in an FET are for the whole temperature 
range higher than the electron mobilities, and both are higher than the mobilities 
determined in an MIS diode. Additionally the temperature dependence of the MIS diode mobilities 
is more pronounced than the temperature dependence of the FET mobilities. The difference
between FETs and MIS diodes in terms of charge carrier density and anisotropy will 
be discussed in detail in the following.

First, we present the activation energies $E_\mathrm{act}$ in Table~\ref{tabenergy}, 
as determined by the Arrhenius behavior of the thermally activated mobility: 
\begin{equation}
\label{eq_ae}
    \mu = \mu_0 \exp\left( - \frac{E_\mathrm{act}}{kT}\right).
\end{equation}

\begin{table}[t]
\renewcommand{\arraystretch}{1.2}
\centering
  \caption{Activation energies for electrons and holes in ambipolar MIS diodes and ambipolar 
   FETs determined by Eq.~(\ref{eq_ae}), as well as barrier energies in FETs determined by 
   Eq.~(\ref{eq_gb}). In addition, the respective fit parameters, $\mu_0$ and $vl$, are given.}
\label{tabenergy}

\begin{tabular}{c cc cc}\hline
Devices & \multicolumn{2}{c}{Electron transport}&\multicolumn{2}{c}{Hole transport}\\ \hline\hline
&$E_\mathrm{act}$ [meV] &$\mu_0$ [cm$^2$/Vs] &$E_\mathrm{act}$ [meV] &$\mu_0$ [cm$^2$/Vs]\\ \hline
MIS diode & 410 & 0.41 & 452 & 0.45 \\
FET       & 183 & 0.18 & 120 & 0.19 \\ \hline
&$E_\mathrm{bar}$ [meV] &$vl$ [cm$^2$/s] &$E_\mathrm{bar}$ [meV] &$vl$ [cm$^2$/s] \\ \hline
FET       & 100 & 56 & 74 & 44 \\ \hline
\end{tabular}
\end{table}

\noindent
Here $\mu_0$ is the high temperature limit of the mobility.
A strong difference in the activation energies is observed 
between the two device types, and also between the two charge carrier types. While for
holes the mobility is higher and the activation energy is lower than for electrons
in the FET, the behavior is reversed for the MIS diode. Note also that the 
activation energies in the MIS diode are more than two times higher than those 
in the FET.

While the transport in the FET is restricted to the accumulation layer with a high charge 
carrier density, the transport in the MIS diode extends over the whole device thickness. 
Thus the charge carrier density in the FET is determined by the effective gate voltage 
($V_\mathrm{G}-V_\mathrm{T}$), and is of the order of $10^{19}$\,cm$^{-3}$. In contrast the 
charge carrier density in the MIS diode is defined by the doping of the material which can be 
determined from the $C$-$V$ measurement. In the MIS diodes analyzed here the charge carrier 
density is two orders of magnitude lower than in an FET. 

In order to describe the dependence of the mobility on the 
charge carrier density extended disorder models were introduced. The extended Gaussian disorder 
model is applicable to polymeric systems \cite{coehoorn05,pasveer05}, while the extended 
correlated disorder model can be used for molecular materials \cite{bouhas09}. 
Both models are reasonable for disordered systems but not for (partially) ordered and 
anisotropic materials \cite{grecu06}. The scanning force microscopy
(SFM) image in Fig.~\ref{figstruct}b shows the 
CuPc film to be polycrystalline; therefore the description using a disorder parameter like the 
width of the Gaussian density of states is not applicable. 
Instead, for polycrystalline films the activation energy for hopping at the grain boundaries 
is an important parameter. Hence the following temperature dependence \cite{Bourguiga07},
  \begin{equation}
  \label{eq_gb}
    \mu = \frac{e v l}{8 k T} \exp\left( - \frac{E_\mathrm{bar}}{kT}\right),
  \end{equation}
was proposed for thermionic emission of charge carriers at grain boundaries, where $v$ is 
the mean charge carrier velocity, and $l$ the grain size. Indeed there is
a distinct difference between path 1 and path 2 which is revealed in the SFM image,
compare Fig.~\ref{figstruct}: Grain boundaries are important for the path parallel
to the semiconductor/insulator interface (path 1), which corresponds to the transport
in the FET accumulation layer, while they are obviously absent for the perpendicular
path (path 2). Hence Eq.~(\ref{eq_gb}) may be applicable for the FET configuration, and
we fit the FET data shown in Fig.~\ref{figmu} also with this formula. The resulting
barrier energies $E_\mathrm{bar}$ are also given in
Table~\ref{tabenergy}. They show the same tendency as the FET activation energies. Note,
however, that in view of the limited temperature range of our data both fits, with 
Eq.~(\ref{eq_ae}) and Eq.~(\ref{eq_gb}), respectively, are of similar quality. At present
the origin of the lower activation or barrier energy for the hole transport relative to 
the electron transport remains unclear. 

In addition to the difference in charge carrier density and film morphology, the anisotropy 
due to the ordered arrangement of the molecules needs to be considered. The CuPc molecules are 
standing almost upright on the surface as schematically illustrated in Fig.~\ref{figstruct}c 
\cite{xrd}. For this reason a good $\pi$-$\pi$ overlap  
is present for the transport parallel to the semiconductor/insulator
interface. Due to the standing molecules and the formation of $\pi$ orbitals on the side of the 
disc-like molecule, the overlap between CuPc molecules standing on top of each other is less 
pronounced. However, no models are available at present which are able to
describe the charge carrier density dependence, the anisotropic behavior 
of the transport, and the transport over grain boundaries in organic films consistently.

\section{Transmission through a Au/CuPc/Au model system}

In this section, we present our results for simple model systems, namely
a single CuPc molecule, as well as two CuPc molecules, which are contacted by two gold
chains, respectively. As mentioned above, we are able to explain the main features of
the experimental results on the basis of these models.

\subsection{Contact resistance}
We first study a single CuPc molecule to which Au chains are attached. In addition to 
the electronic structure, we determine the transmission coefficient and hence the 
conductance, from which we are able to draw conclusions about the contact resistance at 
the metal-molecule interface.

Several methods based on electronic structure calculations have been developed to address 
the problem of transmission through nano-contacts.
In particular, two-terminal contact measurements are treated. (Note that,
in contrast to the experimental situation, we do not apply a gate voltage in our studies.)
Most computational methods rely on a combination of DFT and a 
scattering theory at the non-equilibrium Green's functions level,
based on the Landauer-B\"uttiker scheme. The metallic leads are connected to a central region, 
in which the scattering takes place. In particular, we employ the TRANSIESTA  
\cite{transiesta} and the SMEAGOL program package \cite{smeagol}.
Both packages rely on the SIESTA code but treat the surface Green's function differently. 
Within SMEAGOL, the $I$-$V$ curve is calculated self-consistently with a voltage dependent 
transmission coeffcient. 

In the Landauer-B\"uttiker formalism, the self-energies of the left (L) and 
right (R) lead are calculated first. As metallic leads are used, the  screening  within the 
leads ensures that the distortion due to the contact-interface region decays within a few 
nanometers. Then the leads are connected to a central region of interest (molecule, 
nano-contact, interface).  Thus an effective description of the central region (C) 
emerges which includes the properties of the leads (L,R).
In equilibrium, the transmission coefficient is given by the retarded Green's function
$G_C$ of the central region and the lead self energies $\Sigma_{L/R}$. With
$\Gamma_{L/R}={\rm i}[\Sigma_{L/R}(E)-\Sigma_{L/R}^{\dagger}(E)]$, one finds \cite{Meir92}
\begin{equation}
\label{eq6}
T(E,V=0)={\rm Tr}[\Gamma_L G_{C}^{\dagger}\Gamma_R G_{C}] \; .
\end{equation}
Then, the conductance is given by \cite{Buttiker85}
$ G= \frac{2e^2}{h}T(E_F) $, where the factor of two accounts for the spin. Here and below
we are using a short-hand notation, omitting, in particular, the spatial variables. 
For details, see, e.g., Ref.~\cite{smeagol}.

Applying an external transport voltage $V$, non-equilibrium Green's functions have to be considered,
for example, the lesser Green's function
\begin{eqnarray}
\label{eq7}
G^<_C(E)&=&iG_C(E)[\Gamma_L(E-eV/2)f(E+eV/2)\nonumber\\
&&+\Gamma_R(E+eV/2)f(E-eV/2)]G^\dagger_C(E) \; ,
\end{eqnarray}
from which the charge density can be obtained. The Fermi function is denoted by $f$.
Furthermore, the charge current is given by \cite{Meir92}
\begin{equation}
\label{eq8}
I(V)={\frac{2e}{h}}{\int{dE\;T(E,V)[f(E+eV/2)-f(E-eV/2)]}} \; ,
\end{equation}
where $T(E,V)$ is the voltage-dependent generalization of Eq.~(\ref{eq6}).

\begin{figure}
\centering{\includegraphics[width=0.45\textwidth,clip]{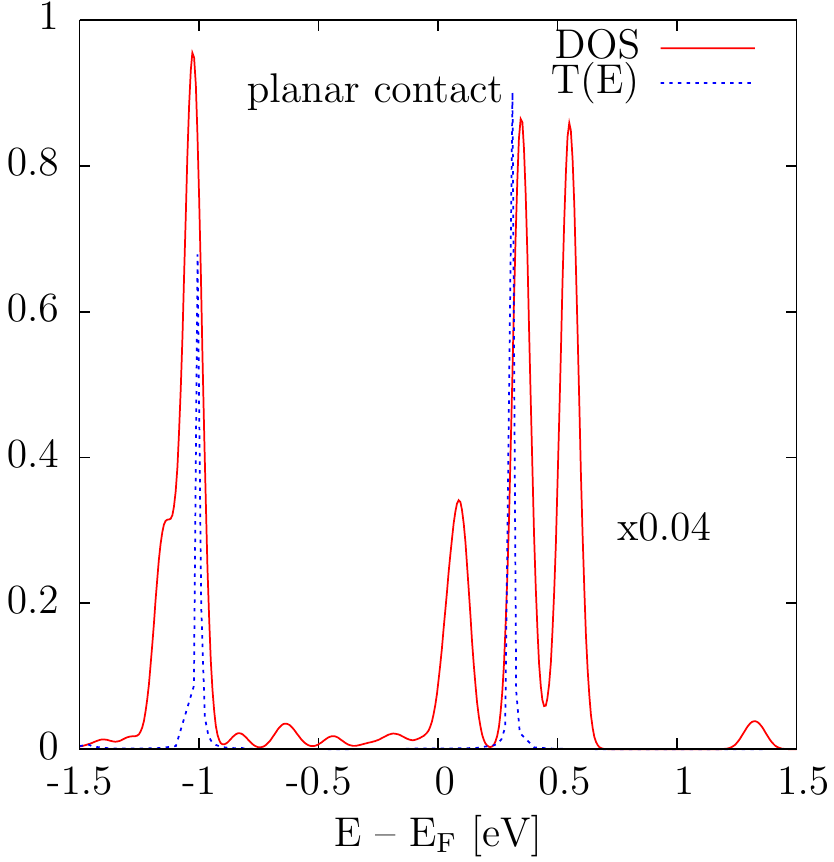}(a)\hfill
\includegraphics[width=0.45\textwidth,clip]{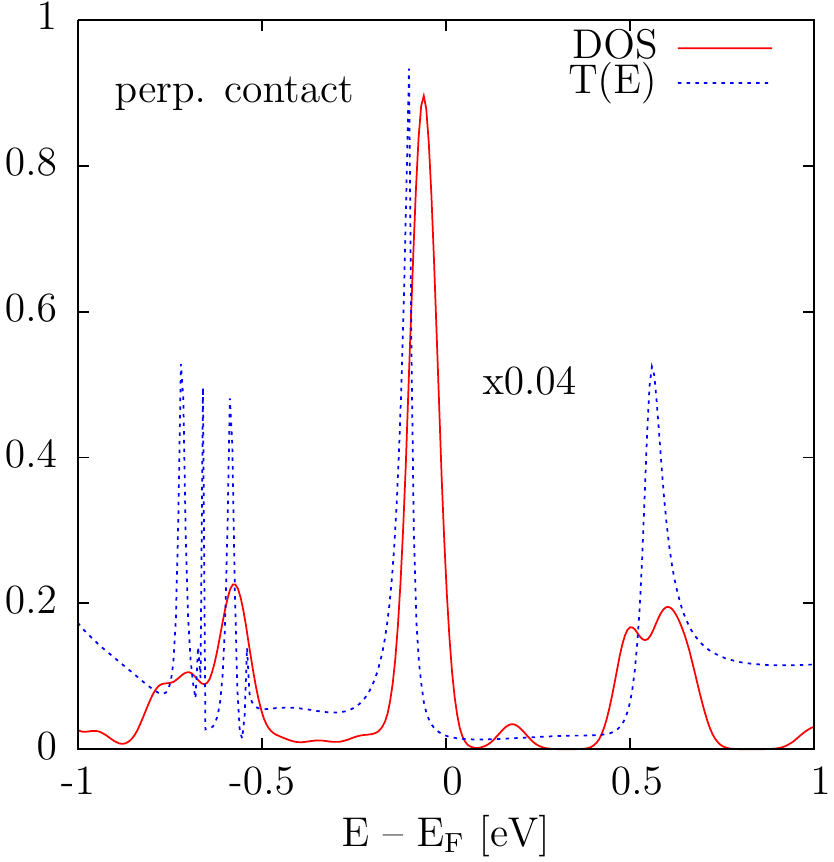}(b)}
\vspace*{0.4cm}\hfill\\
\centering{\includegraphics[width=0.45\textwidth]{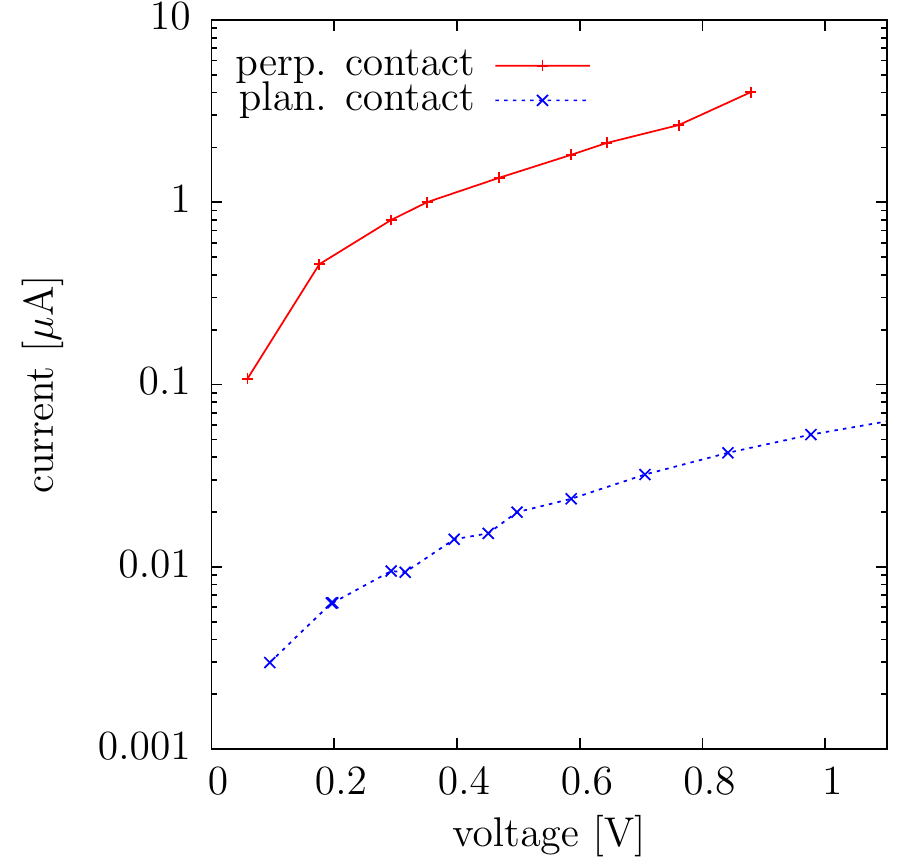}(c)}
\caption{
(a) DOS projected on the CuPc states and $T(E) \equiv T(E,0)$ of a single CuPc molecule 
in the planar contact geometry.
(b) same, in the perpendicular contact geometry.
(c) $I$-$V$ characteristics of both contacts.}
\label{fig_tranp}
\end{figure}

Having discussed above the electronic structure of CuPc attached to Au chains in great detail, 
we turn now to the transport properties. Concerning the planar contact geometry,
the transmission coefficient does not show a resonance at the Fermi level, in contrast
to the DOS projected on the CuPc states, see Figs.~\ref{fig_tranp}a and \ref{fig_tranp}b.
To allow for an easy comparison of DOS and $T(E)$, we scaled 
the DOS by 0.04 (and omitted the units).
As the molecular b$_{1g}$ state at the Fermi level has no spatial overlap with the leads,
it does not contribute to the transmission. 
At the resonant  molecular energy levels, given by the a$_{1u}$ and e$_g$ molecular orbitals 
at $-1.1$ eV and 0.3 eV, the transmission coefficient tends to one. At
the Fermi energy it is given by $T_{\rm par}(E_F)=0.94\cdot 10^{-3}$.

The off-resonance transmission coefficient of the perpendicular contact geometry
clearly is larger than in the planar configuration, see Fig.~\ref{fig_tranp}b.
The states related to the molecular a$_{1u}$ orbital are found predominantly at $-0.2$ eV in the DOS,
but they extend to the Fermi level.
The bonding and anti-bonding b$_{1g}$ states which have a spatial overlap with the Au chains
are located at $\pm 0.6$ eV. Thus we find an energy difference of 1.2 eV between the 
transmitting orbitals along the Au-Cu-Au path. However, in the perpendicular
geometry the molecular orbital near E$_{\rm F}$ also opens a transport channel 
via an Au-N-Au path. Due to this additional channel, the conductance is more than one order 
of magnitude larger for the perpendicular contact than for the planar case.
The perpendicular transmission coefficient at the Fermi energy is given by 
$T_{\rm perp}(E_F)=0.037 \approx 39 T_{\rm par}(E_F)$.

The $I$-$V$ characteristics for the planar and the perpendicular contact are shown 
in Fig.~\ref{fig_tranp}c. For the planar contact, we find an approximately linear 
$I$-$V$ characteristic at low voltage. With
increasing voltage, the transmission becomes resonant and the $I$-$V$ dependence non-linear,
$I\propto V^{3/2}$. The  {$I$-$V$} characteristic of the perpendicular contact can be described 
by $I\propto V^{1.28}$, up to a voltage of $\approx$~1~V. Hence, the current through 
the perpendicular contact is predicted to be about one to two orders of magnitude larger
than in the planar one. 

\subsection{Intermolecular transport}
\begin{figure}
\centering{\includegraphics[width=0.5\textwidth]{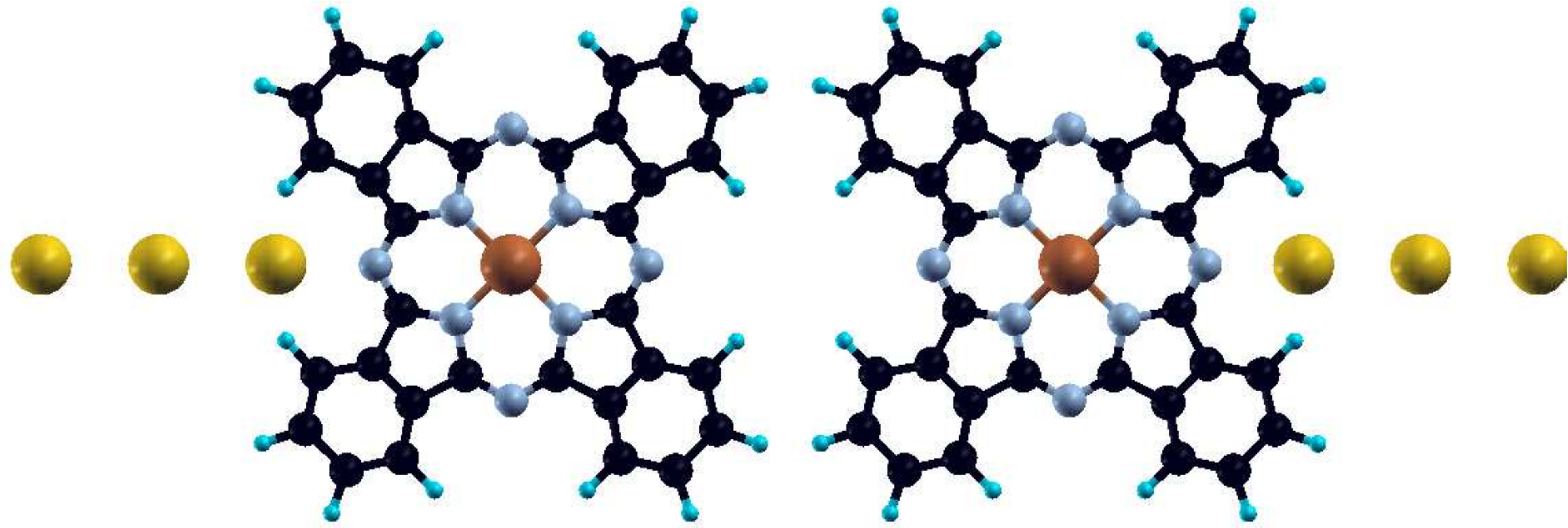}\hspace*{1cm}
 \includegraphics[width=0.30\textwidth,angle=90]{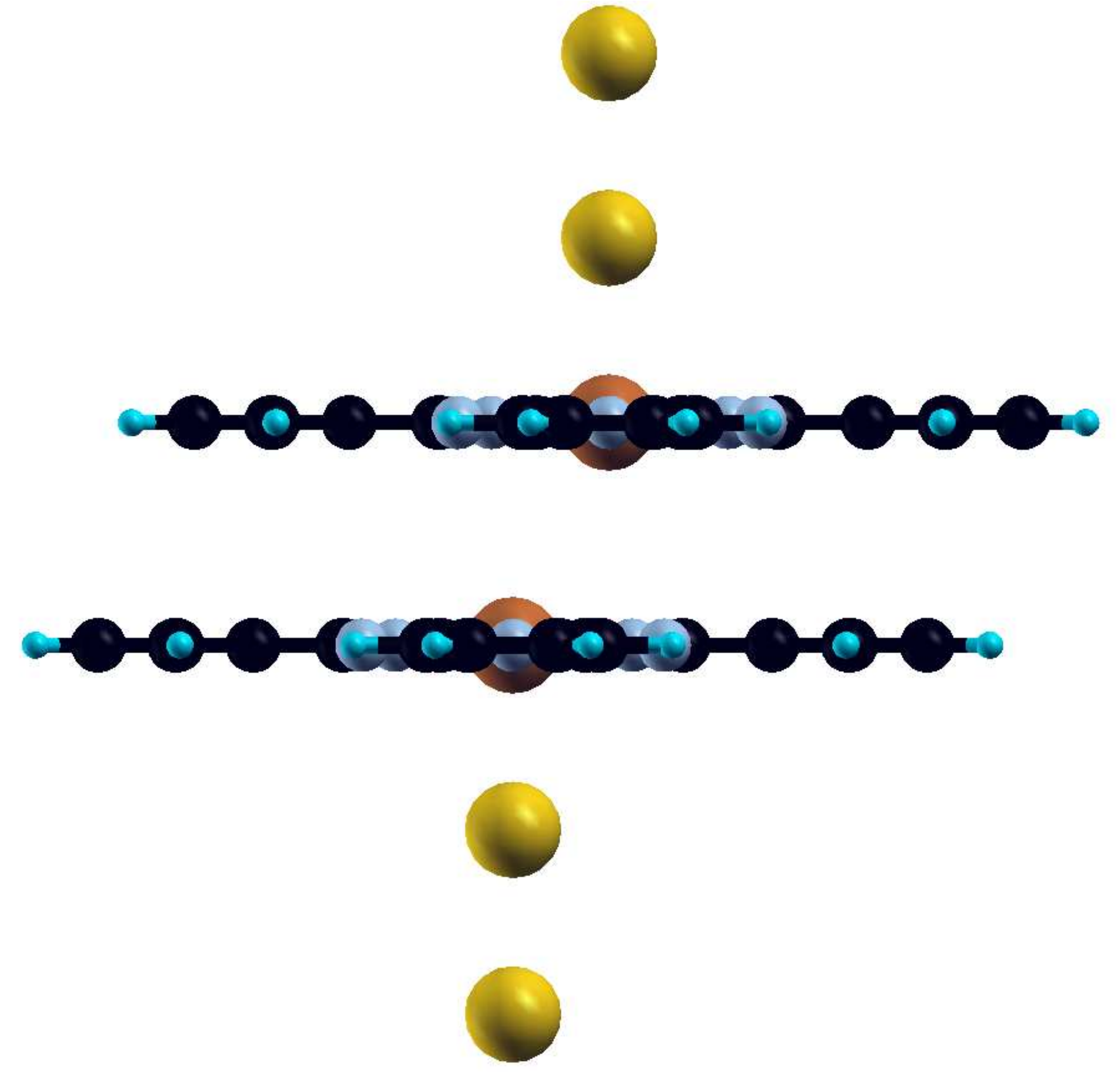}}
  \caption{Left: View from above onto the planar contact with two molecules in the 
   scattering region. Right:
  Sideview of the perpendicular contact. The Cu-Cu distances
  $d=13.72$ \r{A} (planar) and $d=3.79$ \r{A} (perpendicular), respectively, are chosen 
  similar to the distances within a CuPc crystal.}
\label{CuPc_flach_struct_2}
\end{figure}

As the transport in the diodes does not only depend on the injection at the contact, 
i.e., on the contact-molecule coupling, but also on the transport from molecule to molecule,
we study this question by considering two molecules within the central region. However,
in order to be able to simulate a crystal-like arrangement, we will also place the last Au
atoms of the chains near the N atoms between the benzene rings, as indicated in 
Fig.~\ref{CuPc_flach_struct_2}, left hand side, even though in STM 
experiments this case is found to be less stable than the contact to H-C-H \cite{nazin03}.
Then the molecules rotate by 45$^\circ$, similar to the
crystal \cite{bialek}; we call this arrangement $\pi/4$ configuration. 
In the perpendicular geometry we arrange the molecules as shown in
Fig.~\ref{CuPc_flach_struct_2}, right hand side.

\begin{figure}
\centering{\includegraphics[width=0.45\textwidth]{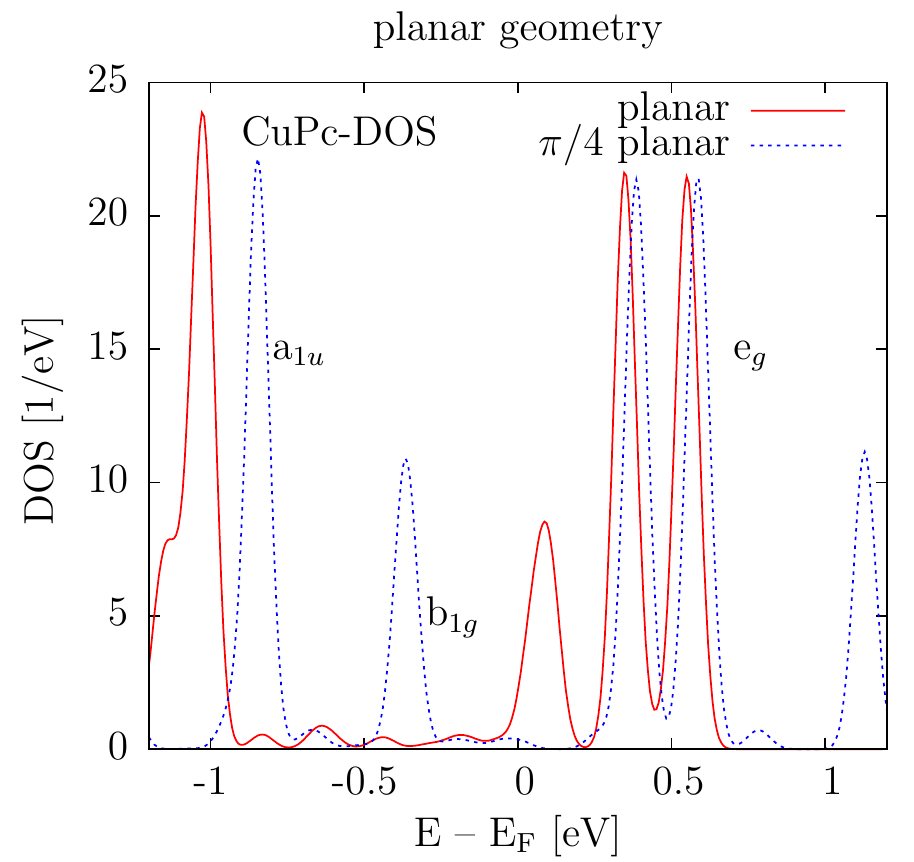}\hfill
\includegraphics[width=0.45\textwidth]{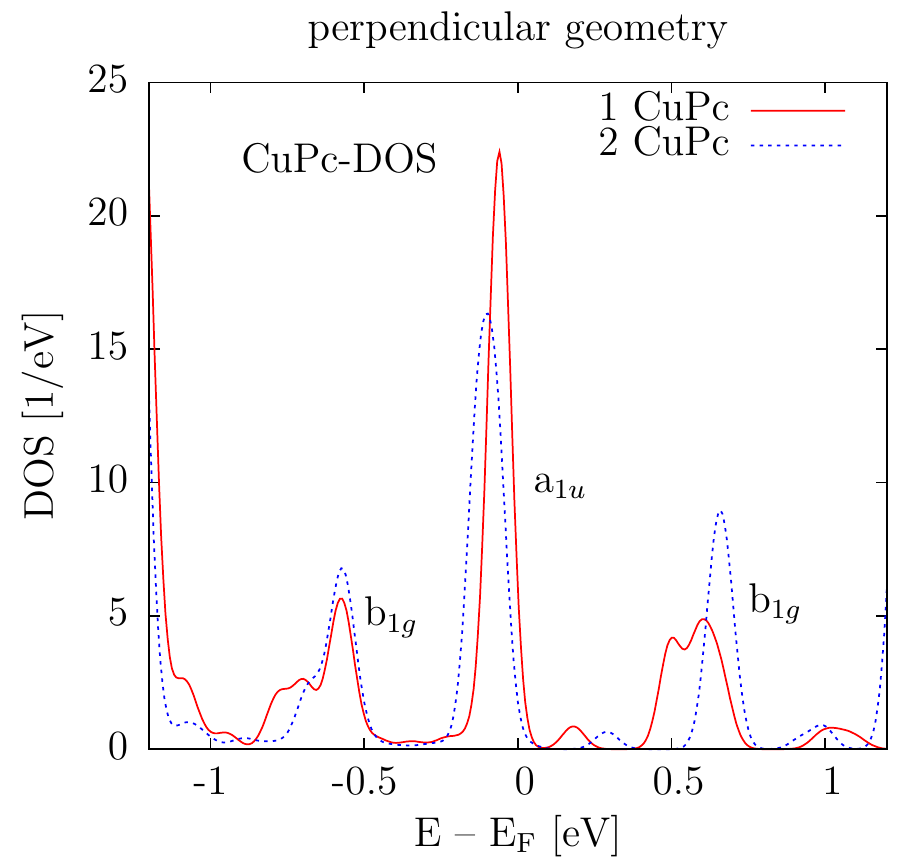}}
 \caption{DOS projected on CuPc. Left: Comparison of the standard and the $\pi/4$
  planar configuration, for a single molecule attached to Au chains.
  Right: Comparison of contacts with one and two molecules, for the
  perpendicular configuration.}
 \label{DOS_12}
\end{figure}

\begin{figure}
\centering{\includegraphics[width=0.45\textwidth,clip]{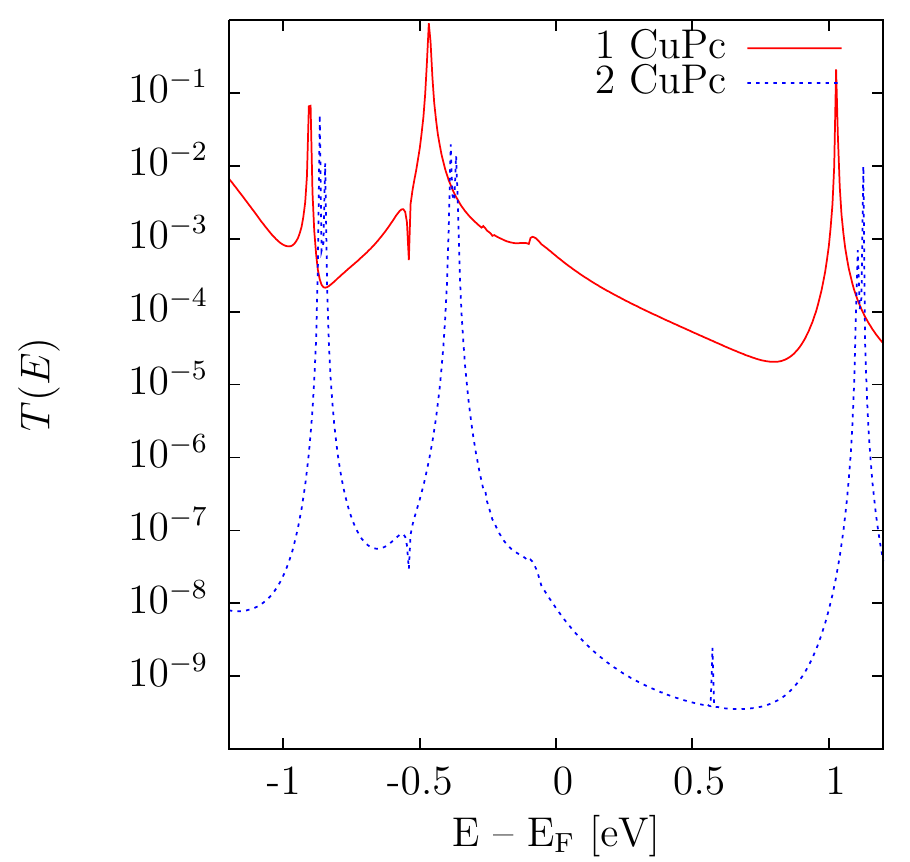}\hfill
 \includegraphics[width=0.45\textwidth,clip]{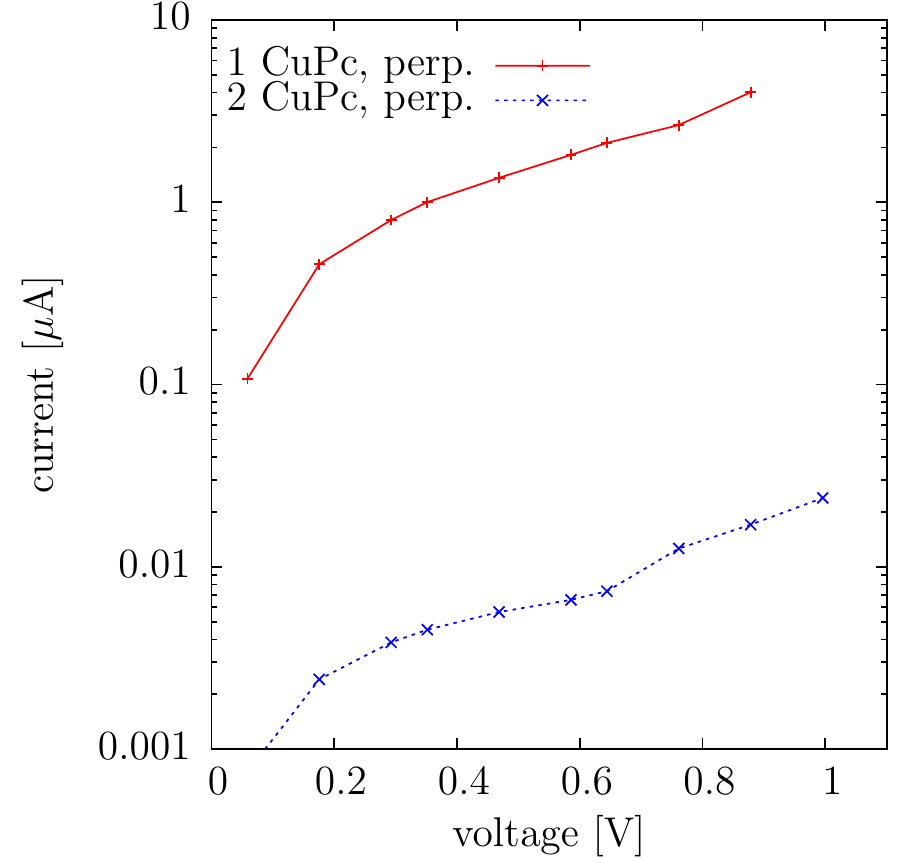}}
\caption{Left: Logarithmic plot of $T(E)$ in the $\pi/4$ planar configuration.  
     Right: Current through one or two molecules in the perpendicular configuration.}
\label{CuPc_12}
\end{figure}

We first compare the electronic structure of the molecular states of the standard and
the $\pi/4$ planar contact for the single-molecule case (with Au chains attached).
In fact there are only minor differences,
see Fig.~\ref{DOS_12}, left panel. The molecular a$_{1u}$ orbital is shifted upwards
to $-0.9$ eV in the $\pi/4$ planar contact. The contribution of the e$_g$ orbital extends 
from 0.3 eV to 0.6 eV in both the standard and the $\pi/4$ contact.
The main difference concerns the Cu b$_{1g}$ orbital which is shifted below the Fermi level, 
to $-0.4$ eV, leading to a nearly insulating contact. Thus the conductance is considerably
smaller for the  $\pi/4$ than for the standard planar contact, even though the transmission 
through the $\pi/4$ planar contact shows a resonant level at the b$_{1g}$ energy,
because a  spatial overlap of Au and Cu is possible in this geometry; compare
Fig.~\ref{CuPc_12}, left hand side.
In all cases, the CuPc DOS of one versus two molecules in the contact region 
is very similar, which is demonstrated for the perpendicular contact 
in Fig.~\ref{DOS_12}, right panel. 

However, the conductance of the planar contact (standard and $\pi/4$) is reduced by a 
factor of 10$^{-5}$ when two molecules are placed within the contact as compared to a single
molecule; the corresponding ratio for a perpendicular contact is only $10^{-2}$, compare
Fig.~\ref{CuPc_12}. Hence
for both contact types, the transport will be determined by the small molecule-molecule
coupling, and we can expect that the ratio $10^{-5}/10^{-2} = 10^{-3}$ also determines
the ratio of mobilities: this assertion roughly agrees with the data, see Table \ref{tabmob}.

We suggest that the three to five orders of magnitude difference between the 
mobilities can also be understood classically by considering the lattice structure. 
(See also the discussion presented in the previous section, last paragraph.)
The mobility in the planar contact configuration 
essentially is determined by hopping processes along the molecules, the distance being
about 14 \AA, while the hopping perpendicular to the molecular plane has to overcome 
a distance of 3 \AA\ only. Thus the large factor of three orders of magnitude 
obtained above indeed appears to be reasonable.

\section{Predictions for fluorinated CuPc}

\begin{figure}
\centering{\includegraphics[width=0.45\textwidth]{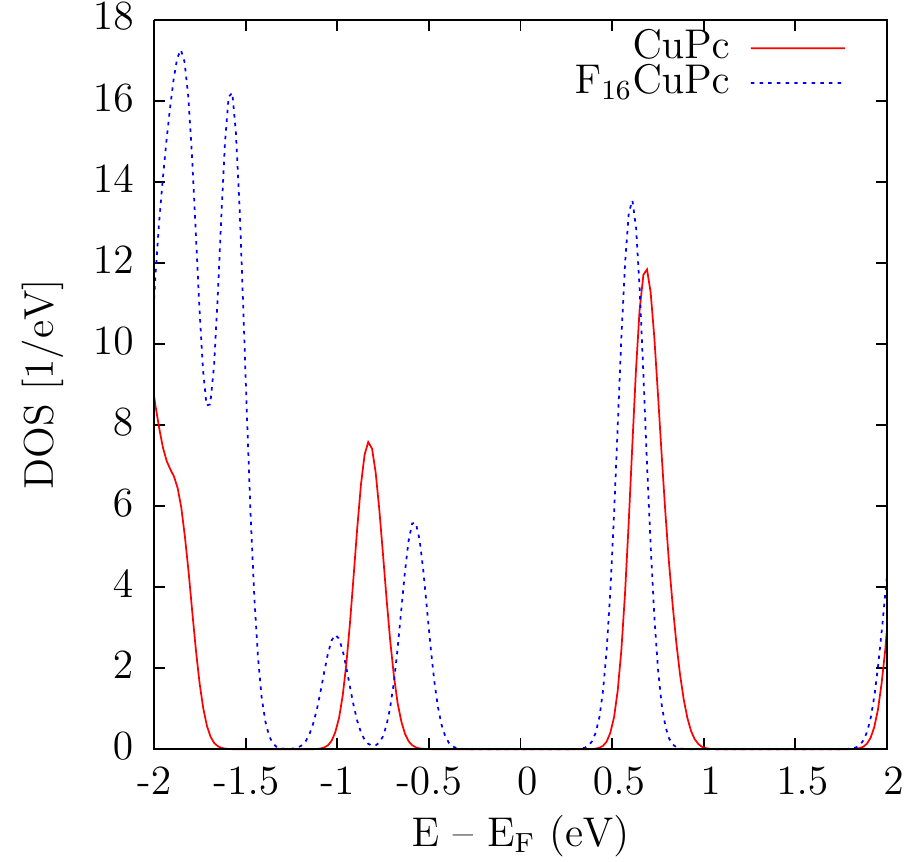}\hfill
\includegraphics[width=0.45\textwidth]{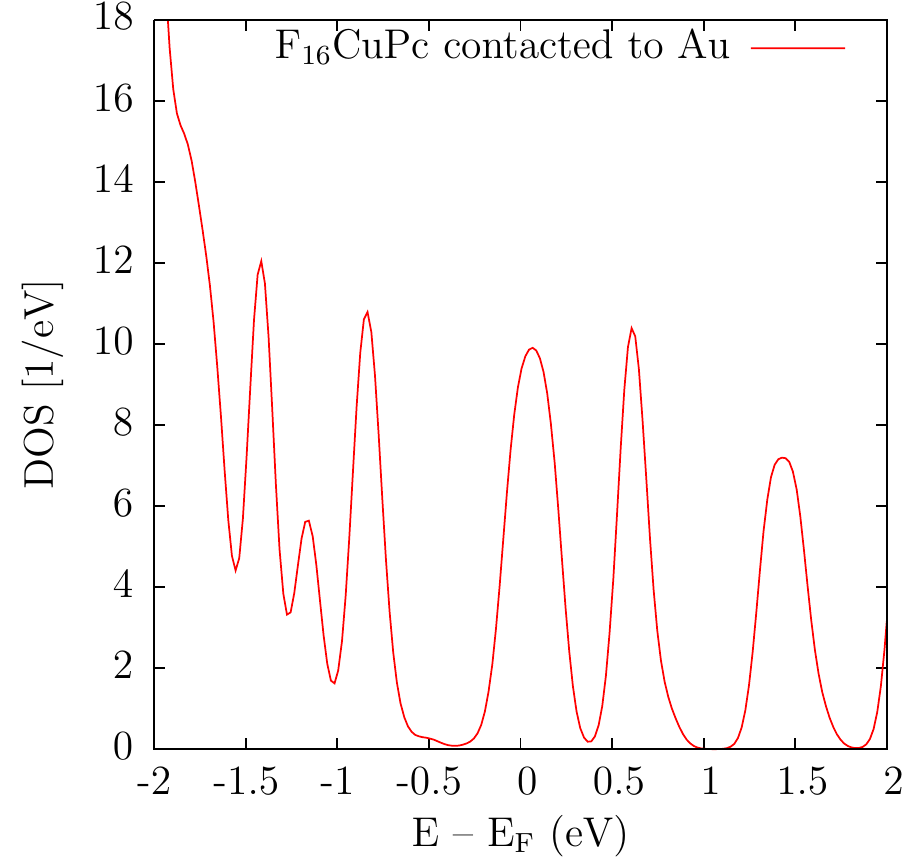}}
\caption{Left: DOS of F$_{16}$CuPc compared to the DOS of CuPc.
   Right: DOS projected on F$_{16}$CuPc, when contacted to 
   Au chains in the planar configuration.}
\label{fig_F}
\end{figure}

Fluorinated F$_{16}$CuPc, where all H atoms are replaced by F atoms, shows a different
behavior. In particular, the Cu b$_{1g}$ state is shifted to lower energy as compared 
to CuPc, and the band gap is reduced; see Fig.~\ref{fig_F}, left hand side.
Our measurements still have to be analyzed in detail, 
but F$_{16}$CuPc appears to show electron transport only \cite{Bao98}.

\begin{figure}
\centering{\includegraphics[width=0.6\textwidth]{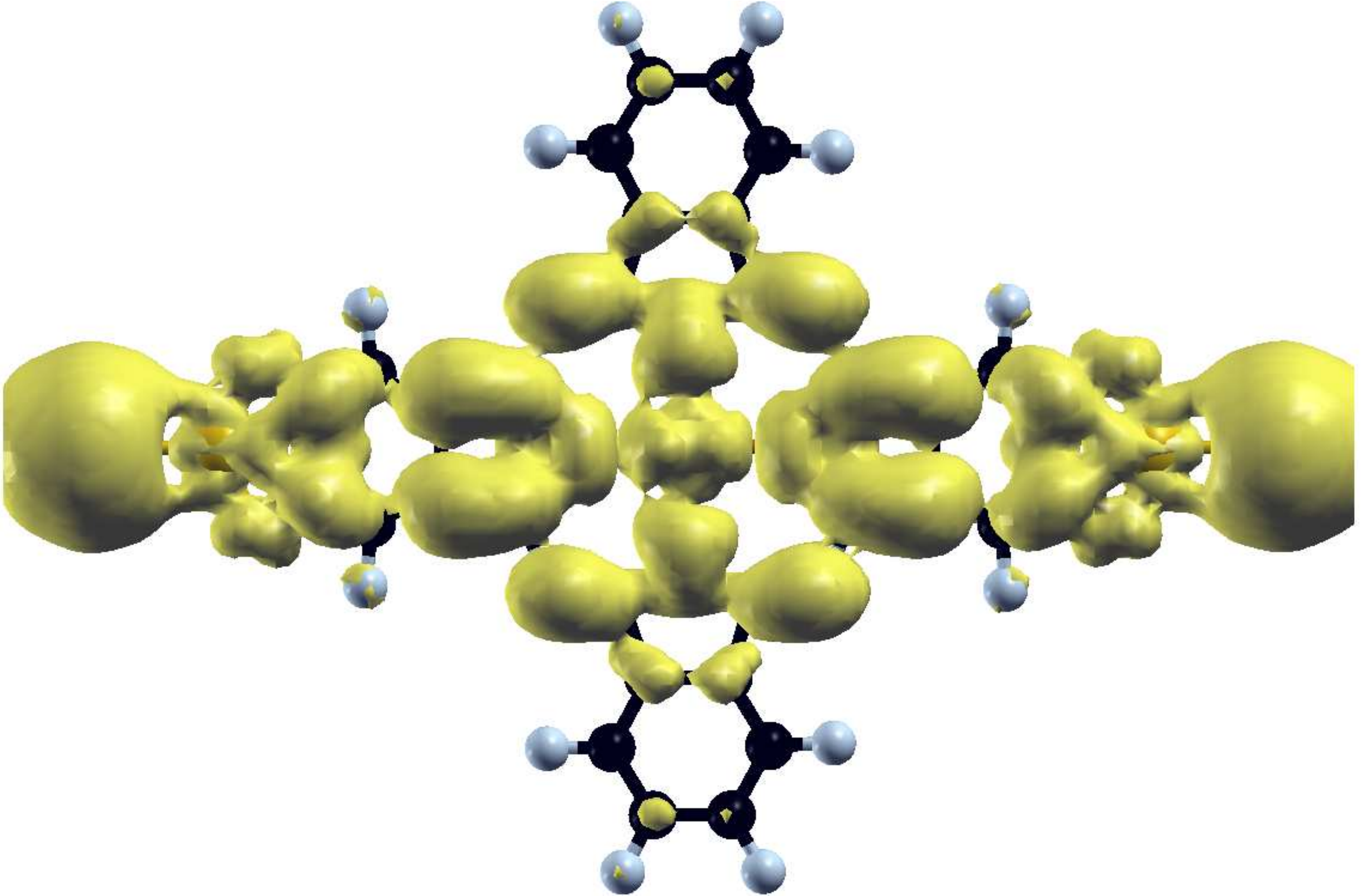}}
\caption{Charge density isosurface of the states at E$_{\rm F}$ for F$_{16}$CuPc contacted 
with Au chains.}
\label{MO_F}
\end{figure}

Within a planar contact geometry, the DOS projected on F$_{16}$CuPc, see Fig.~\ref{fig_F},
right hand side,
shows a peak structure different from the isolated molecule and different from CuPc in the
same geometry. Concerning the charge density isosurface corresponding to the states 
contributing to the symmetric peak at  E$_{\rm F}$, it is not spread over the whole
molecule as in the case of CuPc but quenched towards the transport axis, see Fig.~\ref{MO_F}.
The shape of the charge density does not resemble the charge density of an isolated 
molecule (in strong contrast to CuPc).

The transmission through F$_{16}$CuPc in the planar contact is expected to be higher 
than through CuPc since the DOS shows states at the Fermi level, and the corresponding charge 
density reveals a strong overlap between the leads and the molecular states.
Considering the DFT results for F$_{16}$CuPc in the perpendicular contact, 
we see that the a$_{1u}$ orbital is shifted further downwards below the Fermi level.
Thus hole transmission through F$_{16}$CuPc should be significantly smaller than
through CuPc in the perpendicular configuration. This agrees with experiment as
mentioned above. On the other hand, electron transport through F$_{16}$CuPc 
is enhanced compared to CuPc in the perpendicular configuration. 

\section{Summary}

In summary, our experiments have demonstrated ambipolar transport in CuPc films, i.e.,
electrons and holes move at the same time. We have identified the two relevant transport
paths, path 1 (field-effect transistor (FET), simulated by the perpendicular contact geometry) and path 2
(metal-insulator-semiconductor (MIS) diode, simulated by the planar contact geometry), compare Fig.~\ref{tabmob}.
The mobility is by roughly five orders of magnitude larger for the FET  than for
the MIS diode geometry. For the FET, the hole mobility is larger than the electron
mobility, while for the MIS diode this behavior is reversed. We have analyzed the
temperature dependence of the mobilities using Eqs.\ (\ref{eq_ae}) and (\ref{eq_gb}),
thereby defining the activation and the barrier energy, respectively.
For the FET, the activation
and the barrier energy for holes are smaller than for electrons, while for the MIS diode
the activation energy for electrons is the smaller quantity. However, the temperature
range of our experimental data does not allow discriminating between the minor difference
in the temperature dependence presented by Eqs.\ (\ref{eq_ae}) and (\ref{eq_gb}).

Since the transport properties of CuPc are hardly affected by the electrode material, we
based our theoretical modeling on the assumption that only intrinsic properties are of
relevance. As model systems, we studied molecular nano-contacts, i.e., one or two 
CuPc molecules attached to metallic leads.
The MIS diode geometry was simulated in a planar contact configuration, where CuPc molecules
have been placed between Au chains. The FET geometry was mimiced similarly in a 
perpendicular configuration. The theoretical modeling of the geometry realized in the FET 
measurments reveals that the preferred charge carriers are indeed holes, in agreement with the 
experiments, because the HOMO of the molecule is shifted upwards towards the Fermi level due to the 
coupling with the Au chains. In contrast, for the parallel contact geometry (MIS diode)
the coupling to the chains shifts the LUMO level downwards
and consequently the favored charge carriers are electrons. Accordingly, in the MIS diode
electrons are found to have a higher mobility than holes.
From these results, the coupling of the electrodes to the molecular orbitals apparently 
is the reason for the asymmetry between the charge carrier types. 

To gain a better understanding of the inter-molecular transport we also
considered contacts with two molecules in the central region. We find that in both
cases, planar and perpendicular, the transport is dominated by the molecule-molecule
coupling. In particular, we find that this coupling is much smaller for the planar
(MIS diode) than for the perpendicular (FET) contact, in agreement with the experimental
results. We have presented arguments which indicate that this observation
may be related, from a classical point of view, to the anisotropic  structure of the molecules, which
leads to very different hopping path lengths in the MIS diode versus the FET. 

Of course, further theoretical studies are needed since it is not straightforward to
relate the results of quantum-mechanical transport calculations to room-temperature
measurements. However, in the latter case, it is also very important to
identify the relevant energy levels and their wavefunctions, as well as their overlap,
in order to gain a microscopic understanding of the hopping processes which are 
responsible for high-temperature transport.

Finally we have presented preliminary results for transport through F$_{16}$CuPc. In this
context, further experimental and theoretical investigations are clearly needed.

\medskip

{\small
C.S.\ and U.E.\ acknowledge helpful discussions with U.\ Schwingenschl\"ogl and I.\ Mertig. 
The Au pseudopotential was provided by X.\ Lopez (ETSF, Palaiseau, France). The Smeagol 
project (SS) is sponsored by the Science Foundation of Ireland. This work was supported by the 
German Research Foundation (DFG) through SFB 484 and TRR 80, and by the 
German Academic Exchange Service (DAAD).
}

\end{document}